# Conformation of self-assembled porphyrin dimers in liposome vesicles by phase-modulation 2D fluorescence spectroscopy

Geoffrey A. Lott[1,#,&], Alejandro Perdomo-Ortiz[2,#], James K. Utterback[1], Julia R. Widom[3], Alán Aspuru-Guzik[2] and Andrew H. Marcus[3,*]

[1] Department of Physics, Oregon Center for Optics, University of Oregon, Eugene, OR 97403

[2] Department of Chemistry and Chemical Biology, Harvard University, Cambridge, MA 02138

[3] Department of Chemistry, Oregon Center for Optics, Institute of Molecular Biology, University of Oregon, Eugene, OR 97403

#These authors contributed equally to this work.
&Current address: Boise Technology, Inc., 5465 E. Terra Linda Way, Nampa, ID 83687
*Corresponding author:
ahmarcus@uoregon.edu Tel.: 541-346-4809
Department of Chemistry, Oregon Center for Optics
University of Oregon
Eugene, OR 97403
Fax: 541-346-4643

*Abstract*

By applying a phase-modulation fluorescence approach to 2D electronic spectroscopy, we studied the conformation-dependent exciton-coupling of a porphyrin dimer embedded in a phospholipid bilayer membrane. Our measurements specify the relative angle and separation between interacting electronic transition dipole moments, and thus provide a detailed characterization of dimer conformation. Phase-modulation 2D fluorescence spectroscopy (PM-2D FS) produces 2D spectra with distinct optical features, similar to those obtained using 2D photon-echo spectroscopy (2D PE). Specifically, we studied magnesium meso tetraphenylporphyrin dimers, which form in the amphiphilic regions of 1,2-distearoyl-sn-glycero-3-phosphocholine liposomes. Comparison between experimental and simulated spectra show that while a wide range of dimer conformations can be inferred by either the linear absorption spectrum or the 2D spectrum alone, consideration of both types of spectra constrains the possible structures to a “T-shaped” geometry. These experiments establish the PM-2D FS method as an effective approach to elucidate chromophore dimer conformation.

\body

The ability to determine three-dimensional structures of macromolecules and macromolecular complexes plays a central role in the fields of molecular biology and material science. Methods to extract structural information from experimental observations such as X-ray crystallography, NMR, and optical spectroscopy are routinely applied to a diverse array of problems, ranging from investigations of biological structure-function relationships to the chemical basis of molecular recognition.

In recent years, two-dimensional optical methods have become well established to reveal incisive information about non-crystalline macromolecular systems - information that is not readily obtainable by conventional linear spectroscopic techniques. 2D optical spectroscopy probes the nanometer-scale couplings between vibrational or electronic transition dipole moments of neighboring chemical groups, similar to the way NMR detects the angstrom-scale couplings between adjacent nuclear spins in molecules (1). For example, 2D IR spectroscopy probes the couplings between local molecular vibrational modes, and has been used to study the structure and dynamics of mixtures of molecular liquids, (2), aqueous solutions of proteins (3), and DNA (4). Similarly, 2D electronic spectroscopy (2D ES) probes correlations of electronic transitions, and has been used to study the mechanisms of energy transfer in multi-chromophore complexes. Such experiments have investigated the details of femtosecond energy transfer in photosynthetic protein-pigment arrays (5-8), conjugated polymers (9), and semiconductors (10, 11).

Following the examples established by 2D NMR and 2D IR, 2D ES holds promise as a general approach for the structural analysis of non-crystalline macromolecular systems, albeit for the nanometer length scales over which electronic couplings occur. It is well known that disubstitution of an organic compound with strongly interacting chromophores can lead to coupling of the electronic states and splitting of the energy levels (12-14). The arrangement of transition dipoles affects both the splitting and the transition intensities, which can be detected spectroscopically. Nevertheless, weak electronic couplings relative to the monomer linewidth often limits conformational analysis by linear spectroscopic methods alone. 2D ES has the advantage that spectral information is spread out along a second energy axis, and can thus provide the information needed to distinguish between different model-dependent interpretations. Several theoretical studies have examined the 2D ES of molecular dimers (15-19), and the exciton-coupled spectra of multi-chromophore light harvesting complexes have been experimentally resolved and analyzed (20-22).

Because of its high information content, 2D ES presents previously undescribed possibilities to extract quantum information from molecular systems, and to determine model Hamiltonian parameters (23). For example, experiments by Hayes and Engel extracted such information for the Fenna-Mathews-Olsen light harvesting complex (24). Recently, it was demonstrated by Brinks et al. that single molecule coherences can be prepared using phased optical pulses and detected using fluorescence (25). The latter experiments exploit the inherent sensitivity of fluorescence, and demonstrate the feasibility to control molecular quantum processes at the single molecule level. Fluorescence-based strategies to 2D ES, such as presented in the current work, could provide a route to extract high purity quantum information from single molecules. It may also be a means to study molecular systems in the ultraviolet regime where background noise due to solvent-induced scattering limits ultrafast experiments.

Here we demonstrate a phase-modulation approach to 2D ES that sensitively detects fluorescence to resolve the exciton coupling in dimers of magnesium meso tetraphenylporphyrin (MgTPP), which are embedded in 1,2-distearoyl-sn-glycero-3-phosphocholine (DSPC) liposomal vesicles. MgTPP is a non-polar molecule that preferentially enters the low dielectric amphiphilic regions of the phospholipid bilayer. At intermediate concentration, MgTPP forms dimers as evidenced by changes in the linear and 2D absorption spectra. Quantitative comparison between our measurements and simulated spectra for a broad distribution of selected conformations, screened by a global optimization procedure, shows that the information contained in linear spectra alone is not sufficient to determine a unique structure. In contrast, the additional information provided by 2D spectra constrains a narrow distribution of conformations, which are specified by the relative separation and orientations of the MgTPP macrocycles.

In our approach, called phase-modulation 2D fluorescence spectroscopy (PM-2D FS), a collinear sequence of four laser pulses is used to excite electronic population (26). The ensuing nonlinear signal is detected by sweeping the relative phases of the excitation pulses at approximately kHz frequencies, and by using lock-in amplification to monitor the spontaneous fluorescence. This technique enables phase-selective detection of fluorescence at sufficiently high frequencies to effectively reduce laboratory $1/f$ noise. Because the PM-2D FS observable depends on nonlinear populations that generate fluorescence, a different combination of nonlinear coherence terms must be considered than those of standard photon-echo 2D ES (referred to hereafter as 2D PE). In 2D PE experiments, the signal - a third-order polarization generated from three non-collinear laser pulses - is detected in transmission. The 2D PE signal depends on the superposition of well known nonlinear absorption and emission processes, called ground-state bleach (GSB), stimulated emission (SE) and excited-state absorption (ESA) (27). Analogous excitation pathways contribute to PM-2D FS. However, the relative signs and weights of contributing terms depend on the fluorescence quantum efficiencies of the excited-state populations. Equivalence between the two methods occurs only when all excited-state populations fluoresce with 100% efficiency (28). Thus, self-quenching of doubly-excited exciton population can give rise to differences between the spectra obtained from the two methods -- differences that may depend, in themselves, on dimer conformation. For the conformations realized in the current study, we find that the PM-2D FS and 2D PE methods produce spectra with characteristic features distinctively different from one another.

## Results and discussion

Monomers of MgTPP have two equivalent perpendicular transition dipole moments contained within the plane of the porphyrin macrocycle (see Fig. 1*B, Inset*). These define the molecular-frame directions of degenerate $Q_x$ and $Q_y$ transitions between ground $|g\rangle$ and lowest lying excited electronic states, $|x\rangle$ and $|y\rangle$. The collective state of two monomers is specified by the tensor product $|ij\rangle$ [$i, j \in \{g, x, y\}$], where the first index is the state of monomer 1 and the second that of monomer 2. When two MgTPP monomers are brought close together, their states can couple through resonant dipole-dipole interactions $V_{kl}$ [$k, l \in \{|ij\rangle\}$] with signs and magnitudes that depend on the dimer conformation. We adopt the convention that a conformation is specified by the monomer center-to-center vector $\vec{R}$, which is oriented relative to molecule 1 according to polar and azimuthal angles $\theta$ and $\phi$, and the relative orientation of molecule 2 is given by the Euler angles $\alpha$ and $\beta$ (see Fig. 1*A*, and details provided in *SI Text*). The effect of the interaction is to create an exciton-coupled nine-level system, with states labeled $|X_n\rangle$, comprised of a single ground state ($n$ = 1), four singly-excited states ($n$ = 2 - 5), and four doubly-excited states ($n$ = 6 - 9). Transitions between states are mediated by the collective dipole moment, $\vec{\mu}_1 + \vec{\mu}_2$, which also depends on the structure of the complex.

In Fig. 1B are shown vertically displaced linear absorption spectra of MgTPP samples prepared in toluene, and 70:1 and 7:1 1,2-distearoyl-sn-glycero-3-phosphocholine (DSPC):MgTPP liposomes. For the 70:1 sample, the lineshape and position of the lowest energy $Q$(0,0) feature, centered at 606 nm, underwent a slight red-shift relative to the toluene sample at 602 nm. For the elevated concentration 7:1 sample, the lineshape broadened, suggesting the presence of a dipole-dipole interaction and exciton splitting between closely associated monomer subunits.

In principle, it is possible to model the linear absorption spectrum in terms of the structural parameters $\vec{R}$, $\alpha$ and $\beta$ that determine the couplings $V_{kl}$ and the collective dipole moments, and which ultimately determine the energies and intensities of the ground-state accessible transitions. To test the sensitivity of the linear absorption spectrum to different conformational models, we numerically generated approximately 1000 representative conformations and simulated their linear spectra (details provided in *SI Text*). By comparing experimental and simulated data, we established that a wide distribution of approximately 100 conformations can reasonably explain the linear absorption spectrum. Nevertheless, only a very small conformational sub-space could be found to agree with the experimental 2D spectra (presented below), and which is also consistent with the linear spectrum. In Fig. 1*C* is shown the simulated linear spectrum and the four underlying component transitions of the optimized "T-shaped" conformation. The linear spectrum corresponding to this conformation is composed of two intense spectral features at 16,283 $cm^{-1}$ and 16,619 $cm^{-1}$, one weak feature at 16,718 $cm^{-1}$, and one effectively dark feature at 16,382 $cm^{-1}$ (see *SI Text* for intensity values). The relatively unrestrictive constraint imposed on dimer conformation by the linear spectrum is a consequence of the many possible arrangements and weights that can be assigned to the four overlapping Gaussian features with broad spectral width.

The PM-2D FS method uses four collinear laser pulses to resonantly excite electronic population, which depends on the overlap between the lowest energy electronic transition [the $Q$(0,0) feature] and the laser pulse spectrum (as shown in Fig. 1*C*). We assigned the nonlinear coherence terms GSB, SE and ESA to time-ordered sequences of laser-induced transitions that produce population on the manifold of singly-excited states ($n$ = 2 - 5) and the manifold of doubly excited states ($n$ = 6 - 9). The theoretically derived expressions for PM-2D FS were found to differ from those of 2D PE (details provided in *SI Text*). This is because ESA pathways that result in population on the doubly-excited states have a tendency to self-quench by, for example, exciton-exciton annihilation or other non-radiative relaxation pathways, so that these terms do not fully contribute to the PM-2D FS signal. In 2D PE experiments, signal contributions to ESA pathways interfere with opposite sign relative to the GSB and SE pathways, i.e. $S^{\mathrm{2D\ PE}} = \mathrm{GSB} + \mathrm{SE} - \mathrm{ESA}$. In PM-2D FS experiments, quenching of doubly-excited state population leads to interference between GSB, SE and surviving ESA pathways with variable relative sign, i.e. $S^{\mathrm{PM\text{-}2D\ FS}} = \mathrm{GSB} + \mathrm{SE} + (1-\Gamma)\mathrm{ESA}$, where $0 \le \Gamma \le 2$ is the mean number of fluorescent photons emitted from doubly-excited states relative to the average number of photons emitted from singly excited states. In our analysis of PM-2D FS spectra (described below), we treated $\Gamma$ as a fitting parameter to obtain the value that best describes our experimental data. As we show below, the difference between signal origins of the two methods can result in 2D spectra with markedly different appearances, depending on the specific dimer conformation.

In Fig. 2 are shown complex-valued experimental PM-2D FS data for the 7:1 lipid:MgTPP sample (top row), the 70:1 lipid:MgTPP (middle row), and the toluene sample (bottom row). Rephasing and non-rephasing data, shown respectively in panels A and B, were processed from independently detected signals according to their unique phase-matching conditions. The two types of spectra provide complementary structural information, since each depends on a different set of nonlinear coherence terms. Both rephasing and non-rephasing 2D spectra corresponding to the 7:1 liposome sample exhibit well resolved peaks and cross-peaks with apparent splitting ~ 340 $\mathrm{cm}^{-1}$. This is in contrast to the 2D spectra obtained from control measurements on the 70:1 liposome and toluene samples, which as expected exhibit only the isolated monomer feature due to the absence of electronic couplings in these samples. The 2D spectra of the 7:1 liposome sample are asymmetrically shaped, with the most prominent features a high energy diagonal peak and a coupling peak directly below it. We note that the general appearance of the 7:1 liposome PM-2D FS spectra is similar to previous model predictions for an exciton-coupled molecular dimer (15-18, 29). We next show that the information contained in these spectra can be used to identify a small sub-space of dimer conformations.

By extending the procedure to simulate linear spectra (described above), we numerically simulated 2D spectra for a broad distribution of conformations (details provided in *SI Text*). We performed a least-square regression analysis that compared simulated and experimental spectra to obtain an optimized conformation consistent with both the 2D and the linear data sets. In our optimization procedure, we treated the fluorescence efficiency $\Gamma$ of doubly-excited excitons as a parameter to find the value that best represents the experimental data. In Fig. 3, we directly compare our experimental and simulated PM-2D FS spectra for the optimized conformation. The values obtained for the parameters of this conformation are $\theta$ = 117.4° , $\phi$ = 225.2°, $\alpha$ = 135.2°, $\beta$ = 137.2°, $R$ = 4.2 Å, and $\Gamma$ = 0.31, with associated trust intervals: -16° < $\Delta\theta$ < 4°, -11° < $\Delta\phi$ < 11°, -11° < $\Delta\alpha$ < 11°, -2° < $\Delta\beta$ < 2°, -0.05 Å < $\Delta R$ < 0.05 Å, and -0.1 < $\Delta\Gamma$ = 0.1 (details provided in *SI Text*). For both rephasing and non-rephasing spectra, the agreement between experiment and theory is very good, with an intense diagonal peak and a weaker coupling peak (below the diagonal) clearly reproduced in the simulation. A notable feature of the experimental 2D spectra is the asymmetric lineshape. A possible explanation for these asymmetries is the existence of distinct interactions between the various exciton states and the membrane environment. The discrepancy between experimental and simulated 2D lineshapes is an indication of a shortfall in the model Hamiltonian, which could be addressed in future experiments that focus on system-bath interactions.

In Fig. 4, we show the results of our calculations for three representative conformations. We compare simulated PM-2D FS spectra (with $\Gamma$ = 0.31 optimized to the data, left column), 2D PE spectra (with $\Gamma$ = 2, second column), and linear spectra (third column). It is evident that dimers with different conformations can produce very similar linear spectra. However, these same structures can be readily distinguished by the combined behaviors of both linear and 2D spectra. We note that for both PM-2D FS and 2D PE methods, the 2D spectrum depends on dimer conformation. However, we found that the qualitative appearance of simulated PM-2D FS spectra appear to vary over a greater range, and to exhibit a higher sensitivity to structural parameters in comparison to simulated 2D PE spectra.

Our confidence in the conformational assignment we have made is quantified by the numerical value of the regression analysis target parameter $\chi_{tot}^2 = \chi_{linear}^2 + \chi_{2D}^2$ = 7.39

+ 9.87 = 17.26, which includes contributions from both linear and 2D spectra. By starting with this conformation and incrementally scanning the structural parameters $\theta$ , $\phi$ , α and β, we observed that $\chi^2_{tot}$ increased, indicating that the favored conformation is a local minimum when both linear and 2D spectra are included in the analysis (see Table S1). Similarly, we found that the value Γ = 0.31 corresponds to a local minimum (see Table S2). If only one of the two types of spectra is included, the restrictions placed on the dimer conformation are significantly relaxed. As shown in Fig. 4, conformations that depart from the optimized structure do not simultaneously produce 2D and linear spectra that agree well with experiment.

We found that the average conformation for the MgTPP dimer is a T-shaped structure with mean separation between Mg centers $R$ = 4.2 Å. Close packing considerations alone would suggest the most stable structure should maximize $\pi - \pi$ stacking interactions. However, entropic contributions to the free energy due to fluctuations of the amphiphilic interior of the phospholipid bilayer must also be taken into account. It is possible that the average conformation observed is the result of the system undergoing rapid exchange amongst a broad distribution of energetically equivalent structures. In such a dynamic situation, the significance of the observed conformation would be unclear. However, at room temperature the DSPC membrane is in its gel phase (30), and static disorder on molecular scales is expected to play a prominent role. It is possible that the observed dimer conformation - an anisotropic structure - is strongly influenced by the shapes and sizes of free volume pockets that form spontaneously inside the amphiphilic membrane domain. Future PM-2D FS experiments that probe the dependence of dimer conformation on temperature and membrane composition could address this issue directly.

We have shown that PM-2D FS can uniquely determine the conformation of a porphyrin dimer embedded in a non-crystalline membrane environment at room temperature. The appearance of the PM-2D FS spectra is generally very different from that produced by simulation of the 2D PE method. This effect is due to partial self-quenching of optical coherence terms that generate population on the manifold of doubly-excited states. In the current study on MgTPP chromophores in DSPC liposomes, we find that PM-2D FS spectra are quite sensitive to dimer conformation (20-22).

The PM-2D FS method might be widely applied to problems of biological and material significance. Spectroscopic studies of macromolecular conformation, based on exciton-coupled labels could be practically employed to extract detailed structural information. Experiments that combine PM-2D FS with circular dichroism should enable experiments that distinguish between enantiomers of chiral structures. PM-2D FS opens previously undescribed possibilities to study exciton-coupling under low light conditions, in part due to its high sensitivity. This feature may facilitate future 2D experiments on single molecules, or UV-absorbing chromophores.

## Methods

**Liposome sample preparation.** Samples with 7:1 and 70:1 DSPC:MgTPP number ratio were prepared according to the procedure described by MacMillan et al. (31). An additional control sample was prepared by dissolving MgTPP in spectroscopic grade toluene. Details are provided in *SI Text*.

**Linear absorption spectra.** All samples were loaded into quartz cuvettes with 3 mm optical path lengths. Concentrations were adjusted so that the optical density was ≃ 0.15 at 602 nm. Absorption spectra for each sample was measured using a Cary 3E spectrophotometer (Varian, resolution < 0.7 nm), over the wavelength range 520 - 640 nm. Each spectrum showed the vibronic progression of the lowest lying electronic singlet transition with $Q(0,0)$ centered at approximately 602 nm in the toluene sample, and $Q$ (0,0) centered at approximately 606 nm in the 70:1 lipid sample. The current work focused on the electronic coupling between monomer $Q(0,0)$ transition dipole moments.

**PM-2D FS.** The PM-2D FS method was described in detail elsewhere (26). Samples were excited by a sequence of four collinear optical pulses with adjustable inter-pulse delays (see *SI Text*). The phases of the pulse electric fields were continuously swept at distinct frequencies using acouto-optic Bragg cells, and separate reference waveforms were constructed from the resultant intensities of pulses 1 and 2, and of pulses 3 and 4. The reference signals oscillated at the difference frequencies of the acousto-optic Bragg cells, which were set to 5 kHz for pulses 1 and 2, and 8 kHz for pulses 3 and 4. The reference signals are sent to a waveform mixer to construct "sum" and "difference" side band references (3 kHz and 13 kHz). These side band references were used to phase-synchronously detect the fluorescence, which isolates the non-rephasing and rephasing population terms, respectively. All measurements were carried out at room temperature. The signals were measured as the delays between pulses 1 and 2, and between pulses 3 and 4 were independently scanned. Fourier transformation of the time-domain interferograms yielded the complex-valued rephasing and non-rephasing 2D spectra. Further details are provided in *SI Text*.

**Computational modeling.** A nonlinear global optimization with 13 variables was performed with the aid of the package KNITRO (32). Five variables define the structural arrangements of the dimer; seven variables are associated with the transition intensities, broadening, and line-shapes for the linear and 2D spectra, and the remaining variable $\Gamma$ accounts for the quantum yield of the doubly-excited manifold relative to the singly-excited manifold. To successfully obtain good simulation/experimental agreement, we designed a nonlinear least-square optimization which included in its target function the six experimental 2D data sets (real, imaginary and absolute value rephasing and non-rephasing spectra) and also a contribution from deviations between the experimental and simulated linear spectra. Further details about the construction of the target function are given in *SI Text*.

**Acknowledgements.** A.H.M. thanks Professor Jeffrey A. Cina of the University of Oregon, and Professor Tadeusz F. Molinski of the University of California at San Diego for useful discussions. This material is based on work supported by grants to A.H.M. from the Oregon Nanoscience and Microtechnologies Institute (N3I Program), and from the National Science Foundation, Chemistry of Life Processes Program (CHE-1105272). A.P.-O. and A.A.-G. were supported as part of the Center for Excitonics, an Energy Frontier Research Center funded by the U.S. Department of Energy, Office of Basic Sciences (DE-SC0001088).

**References**

1. Ernst RR, G. Bodenhausen, and A. Wokaun (1990) *Principles of Nuclear Magnetic Resonance in One and Two Dimensions* (Oxford University Press, Oxford, U.K.).
2. Kwak K, S. Park, M. D. Fayer (2007) Dynamics around solutes and solute-solvent complexes in mixed solvents. *Proc. Nat. Acad. Sci.* 104:14221-14226.
3. Fang C, J. D. Bauman, K. Das, A. Remorino, E. Arnold, R. M. Hochstrasser (2008) Two-dimensional infrared spectra reveal relaxation of the nonnucleoside inhibitor TMC278 complexed with HIV-1 reverse transcriptase. *Proc. Nat. Acad. Sci.* 105:1472-1477.
4. Szyc L, M. Yang, E. T. J. Nibbering, T. Elsaesser (2010) Ultrafast vibrational dynamics and local interactions of hydrated DNA. *Angew. Chem. Int. Ed.* 49:3598-3610.
5. Brixner T, J. Stenger, H. M. Vaswani, M. Cho, R. E. Blankenship, and G. R. Fleming (2005) Two-dimensional spectroscopy of electronic couplings in photosynthesis. *Nature* 434:625-628.
6. Collini E, C. Y. Wong, K. E. Wilk, P. M. G. Curmi, P. Brumer, G. D. Scholes (2010) Coherently wired light-harvesting in photosynthetic marine algae at ambient temperature. *Nature* 463:644-647.
7. Abramavicius D, B. Palmieri, D. V. Voronine, F. Šanda, and S. Mukamel (2009) Coherent multidimensional optical spectroscopy of excitons in molecular aggregates; Quasiparticle versus supermolecule perspectives. *Chem. Rev.* 109:2350-2408.
8. Ginsberg NS, Y.-C. Cheng, and G. R. Fleming (2009) Two-dimensional electronic spectroscopy of molecular aggregates. *Acc. Chem. Res.* 42:1352-1363.
9. Collini E, G. D. Scholes (2009) Coherent intrachain energy migration in a conjugated polymer at room temperature. *Science* 323:369-373.

10. Zhang T, I. Kuznetsova, T. Meier, X. Li, R. P. Mirin, P. Thomas, and S. T. Cundiff (2007) Polarization-dependent optical 2D Fourier transform spectroscopy of semiconductors. *Proc. Nat. Acad. Sci.* 104(36):14227-14232.

11. Stone KW, K. Gundogdu, D. B. Turner, X. Li, S. T. Cundiff, K. A. Nelson (2009) Two-quantum two-dimensional Fourier transform electronic spectroscopy of biexcitons in GaAs quantum wells. *Science* 324:1169-1173.

12. Koolhaas MHC, G. van der Zwan, F. van Mourik, and R. van Grondelle (1997) Spectroscopy and structure of bacteriochlorophyll dimers. I. Structural consequences of nonconservative circular dichroism spectra. *Biophys. J.* 72:1828-1841.

13. Stomphorst RG, R. B. M. Koehorst, G. van der Zwan, B. Benthem, and T. J. Schaafsma (1999) Excitonic interactions in covalently linked porphyrin dimers with rotational freedom. *J. Porphyrins and Phthalocyanines* 3:346-354.

14. Matile S, N. Berova, K. Nakanishi, J. Fleischhauer, and R. W. Woody (1996) Structural studies by exciton coupled circular dichroism over a large distance: Porphyrin derivatives of steroids, dimeric steroids, and Brevetoxin B⊥. *J. Am. Chem. Soc.* 118:5198-5206.

15. Voronine DV, D. Abramavicius, and S. Mukamel (2006) Coherent control of cross-peaks in chirality-induced two-dimensional optical signals in excitons. *J. Chem. Phys.* 125:224504.

16. Szöcs V, T. Pálszegi, V. Lukeš, J. Sperling, F. Milota, W. Jakubetz, and H. F. Kauffmann (2006) Two-dimensional electronic spectra of symmetric dimers: Intermolecular coupling and conformational states. *J. Chem. Phys.* 124:124511.

17. Cho M, and G. R. Fleming (2005) The integrated photon echo and solvation dynamics. II. Peak shifts and two-dimensional photon echo of a coupled chromophore system. *J. Chem. Phys.* 123:114506.

18. Kjellberg P, B. Brüggemann, and T. Pullerits (2006) Two-dimensional electronic spectroscopy of an excitonically coupled dimer. *Phys. Rev. B* 74:024303.

19. Biggs JD, and J. A. Cina (2009) Using wave-packet interferometry to monitor the external vibrational control of electronic excitation transfer. *J. Chem. Phys.* 131:224101.

20. Read EL, G. S. Schlau-Cohen, G. S. Engel, J. Wen, R. E. Blankenship, and G. R. Fleming (2008) Visualization of excitonic structure in the Fenna-Mathews-Olson photosynthetic complex by polarization-dependent two-dimensional electronic spectroscopy. *Biophys. J.* 95:847-856.

21. Read EL, G. S. Engel, T. R. Calhoun, T. Mančal, T. K. Ahn, R. E. Blankenship, and G. R. Fleming (2007) Cross-peak-specific two-dimensional electronic spectroscopy. *Proc. Nat. Acad. Sci.* 104:14203-14208.

22. Schlau-Cohen G, T. R. Calhoun, N. S. Ginsberg, M. Ballottari, R. Bassi, and G. R. Fleming (2010) Spectroscopic elucidation of uncoupled transition energies in the major photosynthetic light-harvesting complex, LHCII. *Proc. Nat. Acad. Sci.* 107:13276-13281.

23. Yuen-Zhou J, A. Aspuru-Guzik (2011) Quantum process tomography of excitonic dimers from two-dimensional electronic spectroscopy. I. General theory and application to homodimers. *J. Chem. Phys.* 134:134505-134501-134519.

24. Hayes D, G. S. Engel (2011) Extracting the excitonic Hamiltonian of the Fenna-Matthews-Olson complex using three-dimensional third-order electronic spectroscopy. *Biophys. J.* 100:2043-2052.

25. Brinks D, F. D. Stefani, F. Kulzer, R. Hildner, T. H. Taminiau, Y. Avlasevich, K. Müllen, N. F. van Hulst (2010) Visualizing and controlling vibrational wave packets of single molecules. *Nature* 465:905-909.

26. Tekavec PF, G. A. Lott, and A. H. Marcus (2007) Fluorescence-detected two-dimensional electronic coherence spectroscopy by acousto-optic phase modulation. *J. Chem. Phys.* 127:214307.

27. Mukamel S (1995) *Principles of Nonlinear Optical Spectroscopy* (Oxford University Press, Oxford).

28. Tan H-S (2008) Theory and phase-cycling scheme selection principles of collinear phase coherent multi-dimensional optical spectroscopy. *J. Chem. Phys.* 129:124501.

29. Knoester J (2002) Optical Properties of Molecular Aggregates. *Proceedings of the International School of Physics "Enrico Fermi" Course*, ed Agranovich M, G. C. La Rocca (IOS Press, Amsterdam), Vol CXLIX, pp 149-186.

30. Zein M, R. Winter (2000) Effect of temperature, pressure and lipid acyl chain length on the structure and phase behaviour of phospholipid–gramicidin bilayers. *Phys. Chem. Chem. Phys.* 2:4545-4551.

31. MacMillan JB, and T. F. Molinski (2004) Long-range stereo-relay: Relative and absolute configuration of 1,n-glycols from circular dichroism of liposomal porphyrin esters. *J. Am. Chem. Soc.* 126:9944-9945.

32. Byrd RH, J. Nocedal, R. A. Waltz (2006) KNITRO: An integrated package for nonlinear optimization. *Large-Scale Nonlinear Optimization*, ed di Pillo G, M. Rorna (Springer-Verlag), pp 35-59.

**Figure Legends**

Figure 1. (A) Energy level diagram of two chemically identical three-level molecules, each with degenerate transition dipole moments directed along the x and y axes of the molecular frames. The inset shows a random configuration of two MgTPP monomers whose relative conformation is defined by the molecular center-to-center vector $\vec{R}$ and the angles $\theta$, $\phi$, $\alpha$ and $\beta$. Electronic interactions results in an exciton-coupled nine-level system, with a single ground state, four non-degenerate singly-excited states, and four doubly-excited states. Multi-pulse excitation can excite transitions between ground, singly-excited, and doubly-excited state manifolds. (B) Absorption spectra of the MgTPP samples studied in this work. Spectra are vertically displaced for clarity. The samples correspond to MgTPP in toluene (bottom), aqueous liposome suspension with 70:1 DSPC:MgTPP (middle), and 7:1 DSPC:MgTPP (top). The dashed vertical line represents the lowest energy monomer transition energy used in our calculations. The insets show molecular formulas for MgTPP and lipid DSPC. (C) Overlay of the 7:1 DSPC:MgTPP absorbance and the laser pulse spectrum. The laser spectrum (solid black curve) has been fit to a Gaussian (dashed gray curve) with center frequency 15,501 $cm^{-1}$ (606 nm), and FWHM = 327.0 $cm^{-1}$ (12 nm). The linear absorbance (solid black curve) is compared to the simulated spectrum (dashed black curve), which is based on the T-shaped conformation shown in the inset. Also shown are the positions of the underlying exciton transitions (discussed in text).

Figure 2. Comparison between rephasing (A) and non-rephasing (B) experimental 2D spectra corresponding to the MgTPP samples of Fig. 1B. Complex-valued spectra are represented as 2D contour plots, with absolute value (left column), real (middle column) and imaginary (right column) parts. The color scale of each plot is linear, and normalized to its maximum intensity feature. Positive and negative contours are shown in black and white, respectively, and are drawn at 0.8, 0.6, 0.4, 0.2, and 0.

Figure 3. Comparison between rephasing (A) and non-rephasing (B) experimental (left columns) and simulated 2D spectra (right columns). Absolute value spectra (top), real part (middle) and imaginary part (bottom). The simulated spectra are based on the optimized T-shaped conformation depicted in Fig. 4 (top row, fourth column) and discussed in the text. Color scale and contours have the same values as in Fig. 2.

Figure 4. Comparison between simulated 2D and linear spectra for three selected dimer conformations. Each simulated linear spectrum (gray dashed curve) is compared to the experimental lineshape for the 7:1 DSPC:MgTPP sample. The laser spectrum is shown fit to a Gaussian (dashed gray curve) with center frequency 15,501 $cm^{-1}$ (606 nm), and FWHM ≃ 327.0 $cm^{-1}$ (12 nm). Also shown are the positions of the underlying exciton transitions. Each of the three conformations produce a linear spectrum in agreement with experiment, while only the first (optimized) conformation produces simulated spectra that agree with PM-2D FS data (with $\Gamma = 0.31$). 2D PE spectra (with $\Gamma = 2$) are shown for comparison. Conformations are shown in the fourth column. The squares indicate the size of the MgTPP molecules, with monomer 1 in blue and monomer 2 in red with their respective $Q_x$ and $Q_y$ transition dipoles indicated. Top row: (optimized) conformation with $\theta = 117.4°$, $\phi = 225.2°$, $\alpha = 135.2°$, $\beta = 137.2°$, $R = 4.2$ Å. Middle row conformation with $\theta = 44.3°$, $\phi = 26.0°$, $\alpha = 29.2°$, $\beta = 138.6°$, $R = 3.7$ Å. Bottom row conformation with $\theta = 82.4°$, $\phi = 18.7°$, $\alpha = 47.9°$, $\beta = 124.0°$, $R = 7.6$ Å. Color scale and contours are the same as in Fig. 2.

Figure 1.

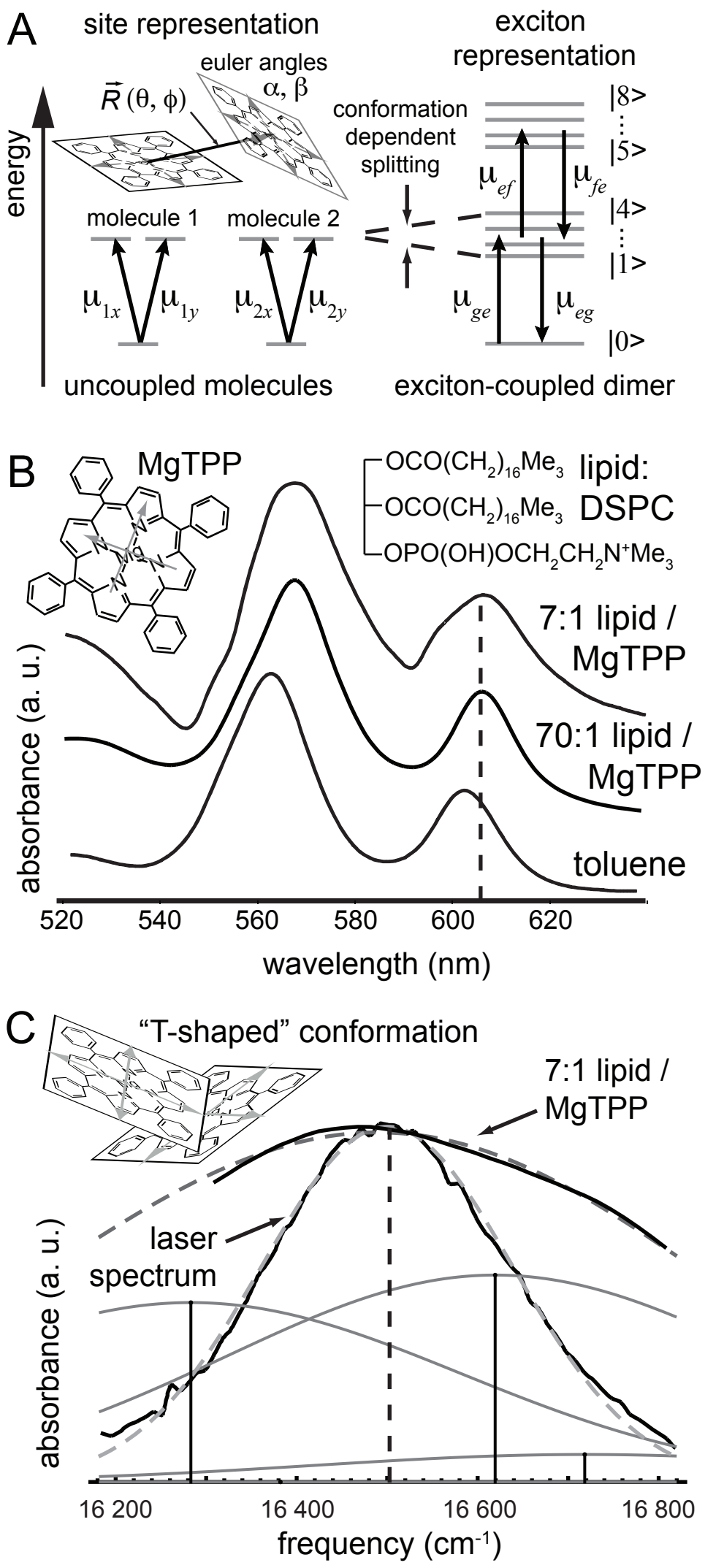

A
site representation
exciton representation
euler angles
α, β
$\vec{R}$ (θ, ϕ)
energy
conformation dependent splitting
|8>
|5>
|4>
|1>
|0>
molecule 1
molecule 2
$\mu_{1x}$
$\mu_{1y}$
$\mu_{2x}$
$\mu_{2y}$
$\mu_{ef}$
$\mu_{fe}$
$\mu_{ge}$
$\mu_{eg}$
uncoupled molecules
exciton-coupled dimer
B
MgTPP
OCO(CH$_2$)$_{16}$Me$_3$
OCO(CH$_2$)$_{16}$Me$_3$
OPO(OH)OCH$_2$CH$_2$N$^+$Me$_3$
lipid: DSPC
7:1 lipid / MgTPP
70:1 lipid / MgTPP
toluene
absorbance (a. u.)
520 540 560 580 600 620
wavelength (nm)
C
"T-shaped" conformation
7:1 lipid / MgTPP
laser spectrum
absorbance (a. u.)
16 200 16 400 16 600 16 800
frequency (cm$^{-1}$)

Figure 2.

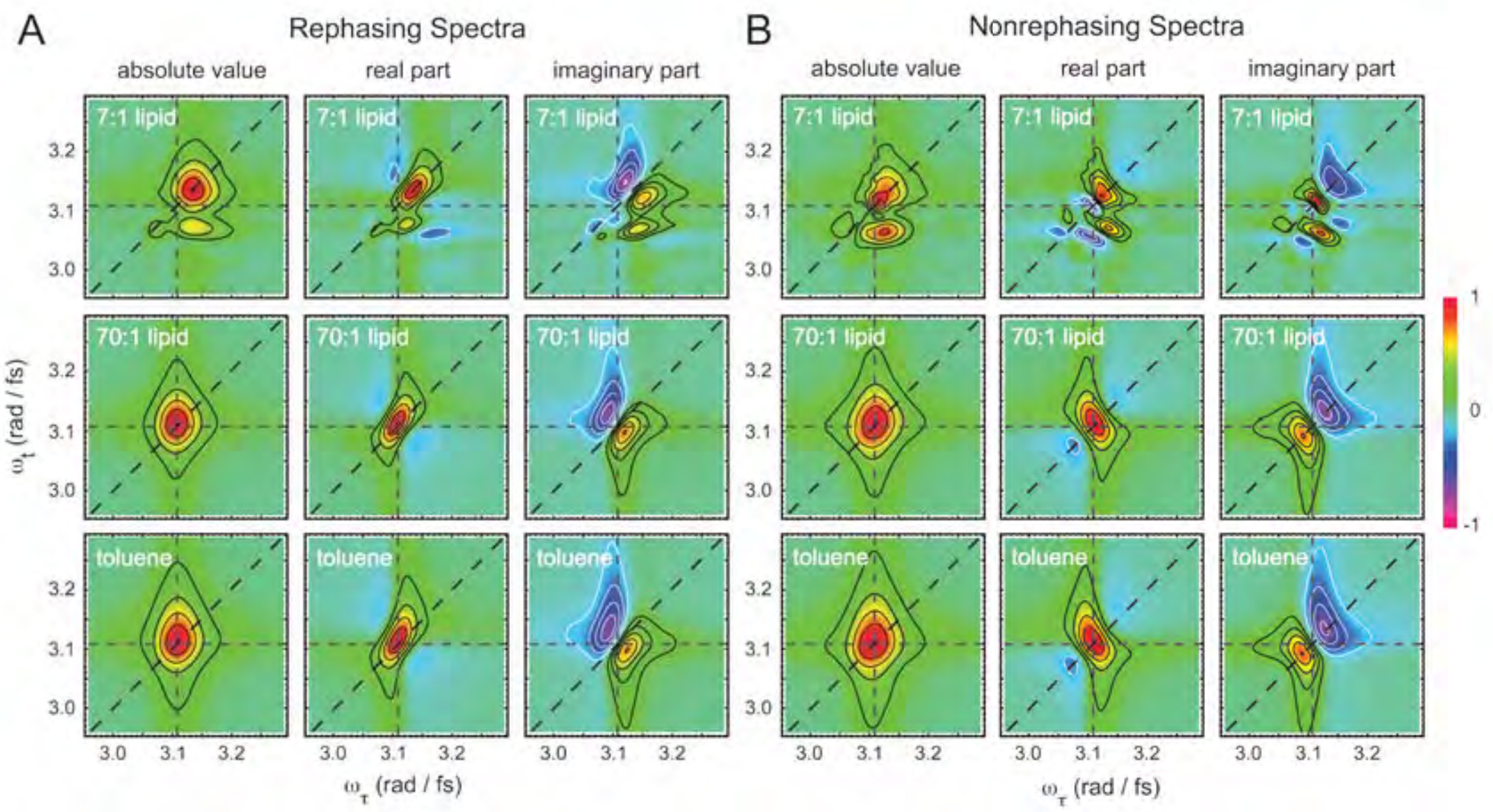

Figure 3.

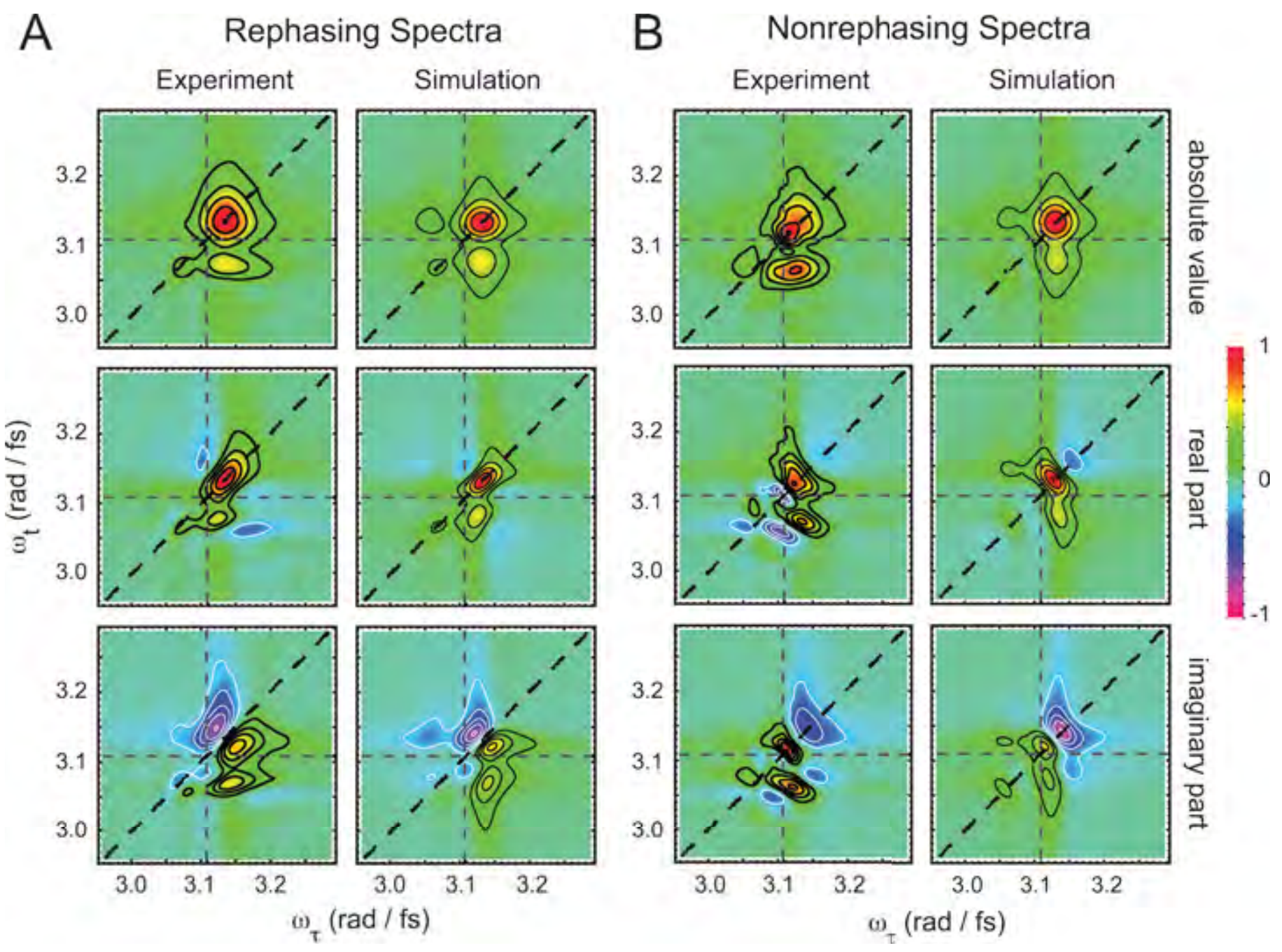

Figure 4.

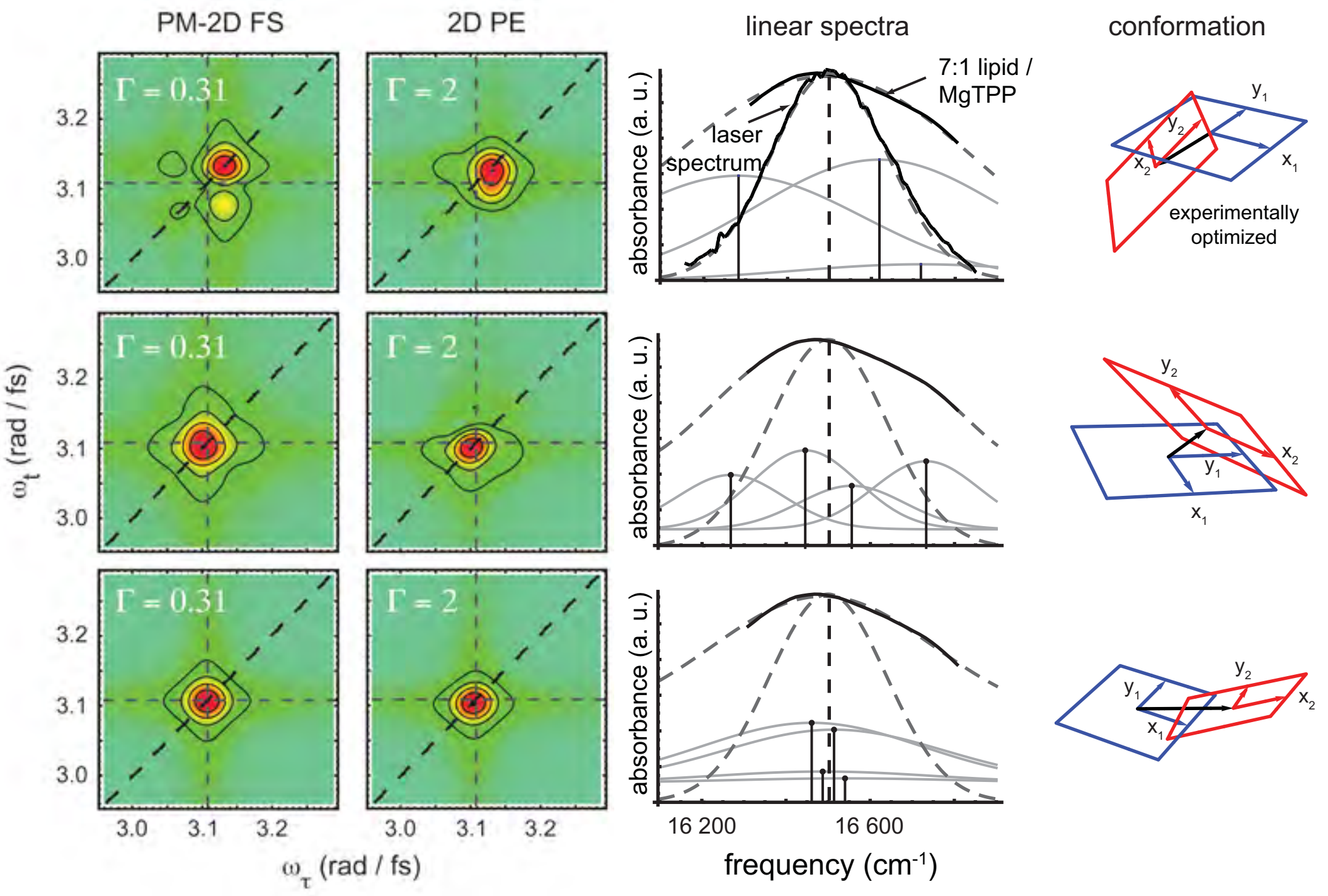

# Supporting Information

## Conformation of self-assembled porphyrin dimers in liposome vesicles by phase-modulation 2D fluorescence spectroscopy

Geoffrey A. Lott[1,#,&], Alejandro Perdomo-Ortiz[2,#], James K. Utterback[1], Julia R. Widom[3], Alán Aspuru-Guzik[2], and Andrew H. Marcus[3,*]

[1] Department of Physics, Oregon Center for Optics, University of Oregon, Eugene, OR 97403

[2] Department of Chemistry and Chemical Biology, Harvard University, Cambridge, MA 02138

[3] Department of Chemistry, Oregon Center for Optics, Institute of Molecular Biology, University of Oregon, Eugene, OR 97403

[#]These authors contributed equally to this work.

[&]Current address: Boise Technology, Inc., 5465 E. Terra Linda Way, Nampa, ID 83687

*Corresponding author:

ahmarcus@uoregon.edu Tel.: 541-346-4809

Department of Chemistry, Oregon Center for Optics

University of Oregon

Eugene, OR 97403

Fax: 541-346-4643

**1. Liposome sample preparation.** Samples were prepared according to the procedure described by MacMillan et al. (1). MgTPP was purchased from Strem Chemicals (Boston), and used without further purification. 1.5 mg of MgTPP was dissolved in 20 mL of toluene, transferred to a 50 ml spherical flask, and the solvent was evaporated. In a separate flask, 12.8 mg of the phospholipid 1,2-distearoyl-sn-glycero-3-phosphocholine (DSPC, Sigma Aldrich) was dissolved in 20 mL of dichloromethane. The contents of the two flasks were combined to create a solution with 7:1 DSPC:MgTPP number ratio. The organic solvent was removed, and 30 ml of nanopure water were added to the flask. The sample was alternately heated to 70˚ C and agitated by ultrasonication for a period of 15 – 30 minutes until an aqueous lipid / porphyrin emulsion was fully formed. The mixture was pre-filtered twice through glass wool, and then extruded through a 100 - 1000 nm pore nylon membrane (Avestin) to create a suspension of liposome vesicles. A second sample with 70:1 DSPC:MgTPP was prepared using the same procedure. It was confirmed using fluorescence microscopy that the MgTPP was localized to the membrane phase. An additional control sample was prepared by dissolving MgTPP in spectroscopic grade toluene.

**2. Phase-modulation 2D Fluorescence Spectroscopy.** The PM-2DFS method was described in detail elsewhere (2). Samples were excited by a sequence of four collinear optical pulses with adjustable inter-pulse delays (see Fig. S1). The pulse sequence was produced using a high repetition regenerative amplifier (Coherent, RegA 9050, 250 kHz, pulse energy ≃ 10 μJ), which was pumped by a Ti:Sapphire seed oscillator (Coherent, Mira, 76 MHz, pulse energy ≃ 9 nJ, pulse width ≃ 35 fs) and a high power continuous wave ND:YVO4 laser (Coherent Verdi V-18, 532 nm). The amplified pulses were sent to two identical optical parametric amplifiers (Coherent, OPA 9400), with output pulse energies ≃ 70 nJ. The relative phase of pulses 1 and 2, and pulses 3 and 4 were independently swept at distinct frequencies (5 kHz and 8 kHz) using acousto-optic Bragg cells. Electronic references were detected from the pulse pairs and sent to a waveform mixer to generate "sum" and "difference" sideband signals (13 kHz and 3 kHz, respectively). These reference waveforms were used to phase-synchronously detect the nonlinear fluorescence, which separately determined the non-rephasing and rephasing signals. The signal phase was calibrated to zero at the origin of the interferograms, i.e. when all inter-pulse delays were set to zero. The measured pulse spectrum at the sample was Gaussian with FWHM ≃ 327 $cm^{-1}$ (≃ 12 nm, shown in Fig. 1*C*). Separate dispersion compensation optics were used for each OPA, and the temporal pulse width determined by autocorrelation was ≃ 60 fs for pulses 1 and 2, and ≃ 80 fs for pulses 3 and 4. The sample cuvette was a flow cell (Starna Cells, 583.3/Q/3/Z15, path length 3 mm, 0.1 mL volume), which was fitted to a peristaltic pump (flow rate ≃ 1 mL / minute, 6 mL reservoir volume). The excitation beam was focused into the sample using a 5 cm focal length lens. Fluorescence from the sample was collected using a 3 cm lens, spectrally filtered (620 nm long-pass, Omega Optical), and detected using an avalanche photo diode (Pacific Silicon Sensor). All measurements were carried out at room temperature. The signals were measured as the delays between pulses 1 and 2, and between pulses 3 and 4 were independently scanned. Fourier transformation of the time-domain interferograms yielded the rephasing and non-rephasing 2D optical spectra.

**3. Exciton-Coupled Dimer of Three-Level Molecules.** Monomers of MgTPP have two equivalent perpendicular transition dipole moments contained within the plane of the macrocycle (see Fig. 1*B, Inset*). These define the directions of degenerate $Q_x$ and $Q_y$ transitions between the ground and lowest lying excited electronic states (3-6). Both transition moments contribute to the collective exciton interactions in a molecular complex, as illustrated in Fig. 1*A*.

To specify dimer conformations, we adopt a molecular-frame coordinate system similar to that described in refs (4) and (5). For each monomer, a right-handed coordinate system is taken with the x and y axes lying parallel to the $Q_x$ and $Q_y$ transition directions, and the z axis perpendicular to the porphyrin plane. We adopt the convention that a conformation is specified by the monomer center-to-center vector $\vec{R}$, which is oriented relative to molecule 1 according to polar and azimuthal angles $\theta$ and $\phi$. The relative orientation of molecule 2 is given by the Euler angles $\alpha$ and $\beta$. Due to the degeneracy of the $Q_x$ and $Q_y$ transitions, all of the results are independent of the third Euler angle, $\gamma$, which we set to zero from this point on (5).

For the Hamiltonian of a dimer of chemically identical three-level molecules in which system-bath effects are neglected, one defines the tensor product states $|ij\rangle$ where *i,j* = *g, x, y* respectively label the states on monomer 1 and 2, and $\{|ij\rangle\}$ is the dimer Hilbert space basis. Notice x (y) is short-hand notation for the excited electronic state associated with the $Q_x$ ($Q_y$) transition on each monomer.

Within this localized basis description, one can write the molecular Hamiltonian for the dimer

$$\underset{\sim}{H} = \underset{\sim}{H}^{(1)} + \underset{\sim}{H}^{(2)} + \underset{\sim}{V} = \underset{\sim}{H}_0 + \underset{\sim}{V}, \tag{S1}$$

where $\underset{\sim}{H}^{(1)}$ ($\underset{\sim}{H}^{(2)}$) is the Hamiltonian associated with monomer 1 (monomer 2). Within the point-dipole approximation, the electronic coupling term can be expressed as $\underset{\sim}{V} = \frac{1}{4\pi\varepsilon R^3}\underset{\sim}{\vec{\mu}}_1 \cdot \left(1 - 3\frac{\vec{R}\vec{R}}{R^2}\right) \cdot \underset{\sim}{\vec{\mu}}_2$, with $\vec{R}$ the monomer center-to-center vector, $\vec{\mu}_1$ ($\vec{\mu}_2$) the dipole operator for monomer 1 (monomer 2), and $\varepsilon$ the dielectric constant.

We simplify our notation by denoting the nine basis states $\{|l_i\rangle\}$, with $|l_1\rangle = |gg\rangle$, $|l_2\rangle = |xg\rangle$, $|l_3\rangle = |yg\rangle$, $|l_4\rangle = |gx\rangle$, $|l_5\rangle = |gy\rangle$, $|l_6\rangle = |xx\rangle$, $|l_7\rangle = |xy\rangle$, $|l_8\rangle = |yx\rangle$, $|l_9\rangle = |yy\rangle$. In this basis, the total Hamiltonian can be written as a nine-by-nine matrix of the form (5):

$$\underset{\sim}{H} \approx \begin{pmatrix} 0 & & & & & & & & \\ & \varepsilon_1 & V_{23} & V_{24} & V_{25} & & & & \\ & V_{32} & \varepsilon_1 & V_{34} & V_{35} & & & & \\ & V_{42} & V_{43} & \varepsilon_1 & V_{45} & & & & \\ & V_{52} & V_{53} & V_{54} & \varepsilon_1 & & & & \\ & & & & & 2\varepsilon_1 & & & \\ & & & & & & 2\varepsilon_1 & & \\ & & & & & & & 2\varepsilon_1 & \\ & & & & & & & & 2\varepsilon_1 \end{pmatrix} \tag{S2}$$

Here we have assumed all the diagonal contributions in the terms associated with $\underset{\sim}{H}_0$, i.e., we have assumed that $\langle l_i | \underset{\sim}{V} | l_i \rangle = 0$ for all $l_i$. To set the reference energy scale, we set $\varepsilon_g^{(i)} = 0$ with $\underset{\sim}{H}^{(i)} | g \rangle = \varepsilon_g^{(i)} | g \rangle$, and therefore $\underset{\sim}{H}_0 | gg \rangle = \left( \varepsilon_g^{(1)} + \varepsilon_g^{(2)} \right) | gg \rangle = 0 | gg \rangle$. The value of $\varepsilon_1$ used in our simulations was 16,500.7 cm$^{-1}$, which corresponds to the monomer excitation energy associated with either of the degenerate $Q_x$ or $Q_y$ transitions for the 70:1 sample (see Fig. 1 in main text). Then $\underset{\sim}{H}_0 | l_k \rangle = \varepsilon_k | l_k \rangle$ with $\varepsilon_k = \varepsilon_1$ for any of the states containing one excitation ($k$ = 2 - 5) and $\varepsilon_k = 2\varepsilon_1$ for the states containing two-excitations ($k$ = 6 - 9). Diagonalization of the Hamiltonian is straightforward since it involves only the $4 \times 4$ block associated with the singly-excited state manifold. Note that the eigen-energies of the singly-excited state manifold correspond to the exciton transitions underlying the region of interest in the experimental and simulated linear spectra. The positions of these eigen-energies depend on the structural parameters of the dimer through the dependence on the couplings:

$$\underset{\sim}{V}_{ij} = \frac{1}{4\pi\varepsilon R^3} \left( \underset{\sim}{\vec{\mu}}_1 \right)_{ij} \cdot \left( 1 - 3\frac{\vec{R}\vec{R}}{R^2} \right) \cdot \left( \underset{\sim}{\vec{\mu}}_2 \right)_{ij} = \frac{|\mu|^2}{4\pi\varepsilon R^3} \kappa_{ij}^2 . \qquad \text{(S3)}$$

Here the orientation factor $\kappa_{ij}^2$ is related to the directions of the transition dipole moments and the vector connecting their centers according to $\kappa_{ij}^2 = \left( \underset{\sim}{\hat{\mu}}_1 \right)_{ij} \cdot \left( \underset{\sim}{\hat{\mu}}_2 \right)_{ij} - \left[ 3 \left( \underset{\sim}{\hat{\mu}}_1 \right)_{ij} \cdot \hat{R} \right] \left[ \hat{R} \cdot \left( \underset{\sim}{\hat{\mu}}_2 \right)_{ij} \right]$, where $\hat{R} = \left( \sin\theta\cos\phi, \sin\theta\sin\phi, \cos\theta \right)$ is the monomer center-to-center unit vector, and $\left( \underset{\sim}{\hat{\mu}}_{\mathrm{n}} \right)_{ij} = \langle \mathrm{l_i} | \underset{\sim}{\vec{\mu}}_{\mathrm{n}} | \mathrm{l_j} \rangle / |\mu|$ is the normalized transition dipole moment operator. The relationship between the square of the monomer transition dipole moment and its absorption coefficient $\alpha$, is given by (7):

$$|\mu|^2 = \frac{3\varepsilon\hbar c}{\pi N_A}\int_{-\infty}^{\infty} d\bar{v}\,\frac{\alpha(\bar{v})}{\bar{v}}. \tag{S4}$$

In Eq. (S4), $\varepsilon$ is the dielectric constant of the medium, $\hbar$ is Planck's constant divided by $2\pi$, $c$ is the speed of light, and $N_A$ is Avogadro's number. The factor $\int_{-\infty}^{\infty} d\bar{v}\alpha(\bar{v})/\bar{v}$ is the optical linewidth of the $Q(0,0)$ transition, measured in wavenumbers, and divided by its peak value. We estimated this number by numerical integration of the lineshape to be 44.3 $M^{-1}$ $cm^{-1}$.

**4. Theoretical Comparison Between PM-2D FS and 2D Photon Echo Spectroscopy (2D PE) Signals.** The PM-2D FS and 2D PE methods are conceptually similar, yet important distinguishing factors can result in their non-equivalence. The 2D PE signal can be interpreted as the third-order polarization of the sample, which is the source of the detected signal field. In contrast, PM-2D FS is a technique based on fluorescence-detection (2). The signal may be considered proportional to the fourth-order excited state population. We thus compare the signals of the two methods based on interpretation of 2D PE signals using third-order perturbation theory, and PM-2D FS signals using fourth-order perturbation theory.

We consider the semiclassical light-matter interaction Hamiltonian,

$$\underset{\sim}{H}_{sc} = \underset{\sim}{H}_0 + \underset{\sim}{H}_{\text{int}}(t), \quad \underset{\sim}{H}_{\text{int}}(t) = -\underset{\sim}{\vec{\mu}} \cdot \vec{\mathrm{E}}(t). \tag{S5}$$

In PM-2D FS experiments, the electric field for $P$ sequential collinear pulses polarized in the $\hat{\mathbf{x}}$ direction can be described by $\vec{\mathrm{E}}(t) = \sum_j^P E_j(t)\hat{\mathrm{x}}$, where

$$E_j(t) = \lambda_j A_j(t - t_j)\cos\left[\omega_j(t - t_j) + \phi_j\right], \tag{S6}$$

with $\lambda_j$ the electric field maximum intensity, $A_j\left(t-t_j\right)=e^{-\frac{4\ln 2}{\tau_{fwhm}^2}\left(t-t_j\right)^2}$ the pulse envelope, and $\omega_j$ is the laser frequency of the $j^{\text{th}}$ pulse. Analogously, in 2D PE experiments the pulses are described by $E_j(t)=\lambda_j A_j\left(t-t_j\right)\cos\left[\omega_j\left(t-t_j\right)-\vec{\mathrm{k}}_j\cdot\vec{\mathrm{r}}\right]$. Using the density matrix formalism, the evolution of the system is described by the Liouville-von Neumann equation

$$i\hbar\frac{\partial\hat{\underset{\sim}{\rho}}(t)}{\partial t}=\left[\hat{\underset{\sim}{H}}_{\text{int}}(t),\hat{\underset{\sim}{\rho}}(t)\right], \tag{S7}$$

where we have used the "hat" notation to indicate that the corresponding operators are in the interaction picture, i.e. $\hat{\underset{\sim}{O}}(t)\equiv e^{i\underset{\sim}{H}_0(t-t_0)}\hat{\underset{\sim}{O}}e^{-i\underset{\sim}{H}_0(t-t_0)}$. A formal solution to Eq. S7 is

$$\hat{\underset{\sim}{\rho}}(t)=\hat{\underset{\sim}{\rho}}(t_0)+\sum_{n=1}^{\infty}\hat{\underset{\sim}{\rho}}^{(n)}(t), \tag{S8}$$

with

$$\begin{aligned}&\hat{\underset{\sim}{\rho}}^{(n)}(t)\equiv\\&(-1)^n\left(\frac{i}{\hbar}\right)^n\int_{t_0}^{t}d\tau_n\int_{t_0}^{\tau_n}d\tau_{n-1}\cdots\int_{t_0}^{\tau_2}d\tau_1\left[\hat{\underset{\sim}{H}}_{\text{int}}(\tau_n),\left[\hat{\underset{\sim}{H}}_{\text{int}}(\tau_{n-1}),\left[\cdots,\left[\hat{\underset{\sim}{H}}_{\text{int}}(\tau_1),\hat{\underset{\sim}{\rho}}(t_0)\right]\cdots\right]\right]\right]\end{aligned} \tag{S9}$$

The expectation of any observable, $\left\langle\hat{\underset{\sim}{O}}(t)\right\rangle\equiv tr\left\{\hat{\underset{\sim}{O}}(t)\hat{\underset{\sim}{\rho}}(t)\right\}$ can be expressed as $\left\langle\hat{\underset{\sim}{O}}(t)\right\rangle=\sum_{n=0}^{\infty}\left\langle\hat{\underset{\sim}{O}}^n(t)\right\rangle$ with $\left\langle\hat{\underset{\sim}{O}}^n(t)\right\rangle\equiv tr\left\{\hat{\underset{\sim}{O}}(t)\hat{\underset{\sim}{\rho}}^{(n)}(t)\right\}$.

As previously mentioned, the 2D PE signal is associated with the third-order polarization and therefore requires

$$\mathbf{P}^{(3)}(t)\equiv tr\left\{\hat{\underset{\sim}{\mu}}(t)\hat{\underset{\sim}{\rho}}^{(3)}(t)\right\}, \tag{S10}$$

while the PM-2D FS signal is associated with the fourth-order excited state population

$$\hat{A}^{(4)}(t) \equiv tr\left\{\hat{\underset{\sim}{A}}(t)\hat{\underset{\sim}{\rho}}^{(4)}(t)\right\}, \tag{S11}$$

with $\underset{\sim}{A} = \sum_\nu |\nu\rangle\langle\nu|$ the projector into all the states $\{|\nu\rangle\}$ of the excited state manifold.

We focus our discussion to the case of the nine-level model of the exciton-coupled dimer (see Fig. 1*A* in the text). 2D PE signals have been derived and studied for this model (8, 9). In Fig. S2, we show the double-sided Feynman diagrams (DSFD) contributing to the non-rephasing and rephasing signals, collected in the phase-matched directions $\mathbf{K}_I \equiv \mathbf{k}_1 - \mathbf{k}_2 + \mathbf{k}_3$ and $\mathbf{K}_{II} \equiv -\mathbf{k}_1 + \mathbf{k}_2 + \mathbf{k}_3$, respectively. Neglecting dissipation for the moment, and assuming the rotating wave approximation in the impulsive limit (8), one obtains the following expressions for each of the non-rephasing terms

$$\mathbf{R}_{1a}^* \propto \sum_{e,e'} \left[\mu_{eg}\mu_{ge}\mu_{e'g}\mu_{ge'}\right]_{\mathbf{e}_1\mathbf{e}_2\mathbf{e}_3\mathbf{e}_4} e^{-i\omega_{eg}\tau} e^{-i\omega_{e'g}t} \tag{S12}$$

$$\mathbf{R}_{2a} \propto \sum_{e,e'} \left[\mu_{eg}\mu_{ge'}\mu_{e'g}\mu_{ge}\right]_{\mathbf{e}_1\mathbf{e}_2\mathbf{e}_3\mathbf{e}_4} e^{-i\omega_{eg}\tau} e^{-i\omega_{ee'}T} e^{-i\omega_{eg}t} \tag{S13}$$

$$\mathbf{R}_{3b}^* \propto \sum_{e,e',f} \left[\mu_{eg}\mu_{ge'}\mu_{e'f}\mu_{fe}\right]_{\mathbf{e}_1\mathbf{e}_2\mathbf{e}_3\mathbf{e}_4} e^{-i\omega_{eg}\tau} e^{-i\omega_{ee'}T} e^{-i\omega_{fe'}t} \tag{S14}$$

Similarly, the rephasing terms are

$$\mathbf{R}_{4a} \propto \sum_{e,e'} \left[\mu_{ge}\mu_{eg}\mu_{ge'}\mu_{ge'}\right]_{\mathbf{e}_1\mathbf{e}_2\mathbf{e}_3\mathbf{e}_4} e^{-i\omega_{ge}\tau} e^{-i\omega_{e'g}t} \tag{S15}$$

$$\mathbf{R}_{3a} \propto \sum_{e,e'} \left[\mu_{ge}\mu_{e'g}\mu_{eg}\mu_{ge'}\right]_{\mathbf{e}_1\mathbf{e}_2\mathbf{e}_3\mathbf{e}_4} e^{-i\omega_{ge}\tau} e^{-i\omega_{e'e}T} e^{-i\omega_{e'g}t} \tag{S16}$$

$$\mathbf{R}_{2b}^* \propto \sum_{e,e',f} \left[\mu_{ge}\mu_{e'g}\mu_{fe'}\mu_{ef}\right]_{\mathbf{e}_1\mathbf{e}_2\mathbf{e}_3\mathbf{e}_4} e^{-i\omega_{ge}\tau} e^{-i\omega_{e'e}T} e^{-i\omega_{fe}t}. \tag{S17}$$

Here, $e, e' \in \{X_2, X_3, X_4, X_5\}$ is the singly-excited state manifold after diagonalization of the 4x4 block of the Hamiltonian in Eq. (S2), $f \in \{X_6, X_7, X_8, X_9\}$ is the doubly-excited state manifold, and $\left[\mu_{ab}\mu_{cd}\mu_{jk}\mu_{lm}\right]_{\mathbf{e}_1\mathbf{e}_2\mathbf{e}_3\mathbf{e}_4}$ denotes the three-dimensional orientationally averaged product $\left\langle(\mu_{ab}\cdot\mathbf{e}_1)(\mu_{cd}\cdot\mathbf{e}_2)(\mu_{jk}\cdot\mathbf{e}_3)(\mu_{lm}\cdot\mathbf{e}_4)\right\rangle$, where $\mathbf{e}_i$ denotes the polarization of the $i^{th}$ pulse (9).

The detailed derivation of these expressions and their relation to the PM-2D FS terms will be published elsewhere. In Fig. S2, we present the corresponding PM-2D FS non-rephasing and rephasing DSFDs obtained from the fourth-order perturbation expansion (Eq. S11). For our current purpose, we provide here the connection to the 2D PE expressions presented in formulas S12 - S17. For example, it can be shown that for the case of the non-rephasing contributions, the following relations between 2D PE and PM-2D FS hold: $\mathbf{R}_{1a}^* = \mathbf{Q}_{5a}^* \equiv \mathrm{GSB}_1$, $\mathbf{R}_{2a} = \mathbf{Q}_{2a} \equiv \mathrm{SE}_1$, $\mathbf{R}_{3b}^* = \mathbf{Q}_{3b}^* \equiv \mathrm{ESA}_1$, and also $\mathbf{Q}_{3b}^* = \mathbf{Q}_{7b}$. For the rephasing signals, we have: $\mathbf{R}_{4a} = \mathbf{Q}_{4a} \equiv \mathrm{GSB}_2$, $\mathbf{R}_{3a} = \mathbf{Q}_{3a} \equiv \mathrm{SE}_2$, $\mathbf{R}_{2b}^* = \mathbf{Q}_{2b}^* \equiv \mathrm{ESA}_2$, and $\mathbf{Q}_{2b}^* = \mathbf{Q}_{8b}^*$.

Although most of the 2D PE and PM-2D FS contributions are equal, there are two key differences that make their signals unique:

1. Since PM-2D FS is a fluorescence-detection technique, it is important to consider the nature of the resulting excited state of the system after the interaction with the four ultrafast pulses. As a consequence, even though mathematically $\mathbf{Q}_{3b}^* = \mathbf{Q}_{7b}$ $\left(\mathbf{Q}_{2b}^* = \mathbf{Q}_{8b}^*\right)$, they do not contribute equally because the terms $\mathbf{Q}_{3b}^*$ $\left(\mathbf{Q}_{2b}^*\right)$ end in the singly-excited manifold $\{|e\rangle\}$ while the terms $\mathbf{Q}_{7b}$ $\left(\mathbf{Q}_{8b}^*\right)$ end in the doubly-excited states $\{|f\rangle\}$. Since the quantum yield of singly- and doubly-excited states are different in general, we must account for this fact when simulating the signals. We introduced a multiplicative factor $\Gamma$ in front of the diagrams ending in a doubly-excited population (see $\mathbf{Q}_{7b}$ and $\mathbf{Q}_{8b}^*$ in Fig.

S2) to capture the relative quantum yield of this doubly-excited state compared to the singly-excited states. Due to the abundance of non-radiative relaxation pathways for highly excited states, one expects the relative quantum yield of the doubly-excited states to be significantly smaller than the singly-excited states. In a fully ideal coherent case, where two-photons are emitted via the pathway $|f\rangle \to |e\rangle \to |g\rangle$, then $\Gamma = 2$. In general, $0 \le \Gamma \le 2$. For the dimer studied in the current work, the value of $\Gamma = 0.31$ was obtained from the global optimization that compared simulated and experimental spectra. A visual illustration of these differences can be found in Fig. 4 of the main text, where we compare for three different conformations PM-2D FS spectra ($\Gamma = 0.31$) to the corresponding 2D PE spectra ($\Gamma = 2$). Table S2 shows the sensitivity of the optimization target function to the parameter $\Gamma$ around the optimal value of 0.31.

2. The GSB, SE and ESA terms add up differently for 2D PE and PM-2D FS. This is a consequence of the third-order versus fourth-order perturbation approach respectively. This is the main reason for the different appearances of PM-2D FS versus 2D PE spectra.

The non-rephasing and rephasing 2D PE signals are written:

$$\begin{aligned} S_{NRP}^{2D\,PE}(\tau,T,t) &\propto \mathbf{R}_{1a}^{*} + \mathbf{R}_{2a} - \mathbf{R}_{3b}^{*} \\ &\propto \mathrm{GSB}_1 + \mathrm{SE}_1 - \mathrm{ESA}_1 \end{aligned} \tag{S18}$$

$$\begin{aligned} S_{RP}^{2D\,PE}(\tau,T,t) &\propto \mathbf{R}_{4a} + \mathbf{R}_{3a} - \mathbf{R}_{2b}^{*} \\ &\propto \mathrm{GSB}_2 + \mathrm{SE}_2 - \mathrm{ESA}_2. \end{aligned} \tag{S19}$$

Taking account of the differences between the two methods mentioned above, and making use of Fig. S2, the non-rephasing and rephasing PM-2D FS signals are written:

$$\begin{aligned} S_{NRP}^{PM-2D\,FS}(\tau,T,t) &\propto -\left(\mathbf{Q}_{5a}^{*} + \mathbf{Q}_{2a} + \mathbf{Q}_{3b}^{*} - \Gamma\mathbf{Q}_{7B}\right) \\ &\propto -\left[\mathrm{GSB}_1 + \mathrm{SE}_1 + (1-\Gamma)\mathrm{ESA}_1\right] \end{aligned} \tag{S20}$$

$$S_{RP}^{PM-2D\,FS}(\tau,T,t) \propto -\left(\mathbf{Q}_{4a}+\mathbf{Q}_{3a}+\mathbf{Q}_{2b}^{*}-\Gamma\mathbf{Q}_{2b}^{*}\right)$$
$$\propto -\left[\mathrm{GSB}_2+\mathrm{SE}_2+(1-\Gamma)\mathrm{ESA}_2\right] \tag{S21}$$

Although the signal expressions corresponding to the two techniques are closely related, the variable sign contribution of the ESA terms in the PM-2D FS expressions (formulas S20 and S21), in comparison to the well known negative sign ESA contribution in 2D PE spectroscopy (formulas S18 and S19), can lead to considerably different appearances of the 2D spectra. The differences in sign assignments of these terms arises from the commutator expansions of Eq. S11.

In the current work, we have considered the case where the population time $T$ = 0 fs. To account for optical dephasing, inhomogeneous broadening and other dissipative processes, we multiplied each term given by Eqs. S18 - S21 by a phenomenological line broadening function, which is assumed to be Gaussian in both coherence times, $\tau$ and t. That is, the rephasing signals were multiplied by the factors $e^{-\tau^2/\sigma_{RP}^2}$ and $e^{-t^2/\kappa_{RP}^2}$. Similarly, we have used factors that contain the parameters $\sigma_{NRP}$ and $\kappa_{NRP}$ to describe the broadening of the non-rephasing signals. Fourier transformation of these equations to the $\omega_\tau$ and $\omega_t$ domains provide the real, imaginary, and absolute value 2D spectra presented in Fig. 3 of the text, with very good agreement to experiment. We note that while the intensities and positions of 2D optical features are well accounted for by the molecular dimer Hamiltonian, the observed spectral lineshapes deviate markedly from this simple model. The asymmetric lineshapes could be due to a number of factors, including differences in the system-bath coupling and population times of the various excited states, as well as the effects of laser pulse overlap. Understanding the origins of the lineshape asymmetries is important to future studies.

**5. Computational Modeling.** The search for the porphyrin-dimer conformation consistent with both linear and 2D experimental data involved a constraint-nonlinear-global optimization with 13 variables. Optimizations performed separately on the linear and 2D spectra did not provide solutions consistent with both sets of experimental data. We therefore employed a joint target optimization function, which involved a least-square regression optimization using both sets of data -- i.e., $\chi^2_{tot} = \chi^2_{lin} + \chi^2_{2D}$, which is described in the next section.

***Construction of target function for linear spectra.*** The $Q(0,0)$ transition of the monomer in the lipid bilayer membrane has energy 16,500.7 cm$^{-1}$ (see 70:1 lipid:MgTPP linear spectra shown in Fig. 1B of the text). The $Q(0,0)$ feature contains contributions from both degenerate $Q_x$ and $Q_y$ transitions. Formation of the electronically coupled dimer results in four new transitions, which arise from the couplings between the states on each monomer. The energies of the resulting exciton transitions are given by the eigenvalues obtained from diagonalization of the $4 \times 4$ block of the Hamiltonian matrix (Eq. S2). The relative intensities of the exciton transitions are computed from the eigenvectors, which determine the transition dipole moments (5). All of the transitions are broadened and modeled as Gaussians centered at their respective eigenvalues, with equal line widths $\sigma_{lin}$. The value of $\sigma_{lin}$ was treated as an optimization parameter. The trial function used to reproduce the linear spectra can be written:

$$\text{trial}_{\text{lin}}\left(\theta,\phi,\alpha,\beta,R,a_0,\eta,\sigma_{\text{lin}}\right) = a_0 + \eta\left\{\sum_{i=1}^{4} a_i\left(\theta,\phi,\alpha,\beta,R\right) e^{-\left[\bar{\nu}-\bar{\nu}_i(\theta,\phi,\alpha,\beta,R)\right]^2/\sigma_{lin}^2}\right\}. \tag{S22}$$

In Eq. S22, $a_0$ accounts for background absorption, η is a multiplicative factor that uniformly adjusts the intensities $a_i$, and $\bar{\nu}_i$ are the eigen-energies of the transitions. All of the optimization parameters are determined by a least-square regression analysis when compared to experimental data. We isolated the experimental data inside the region-of-interest frequency window 16,300 cm$^{-1}$ - 16,810 cm$^{-1}$, which is centered around the uncoupled monomer transition energy ($\varepsilon_1$ = 16,500.7 cm$^{-1}$). We denote the least-square sum as target$_{\text{lin}}$, and the contribution to the total optimization function is defined as $\chi^2_{lin}$ = $10^5$ target$_{\text{lin}}$. For example, the value of $\chi^2_{lin}$ corresponding to the best fit to both linear and 2D spectra is 7.39. The values of the eigen-energies for the optimized conformation are $\bar{\nu}_1$ = 16,283 cm$^{-1}$, $\bar{\nu}_2$ = 16,382 cm$^{-1}$, $\bar{\nu}_3$ = 16,619

cm$^{-1}$, and $\bar{\nu}_4$ = 16,718 cm$^{-1}$, with respective relative intensities $a_1$ = 0.867, $a_2$ = 1.94 × 10$^{-13}$, $a_3$ = 1.00, and $a_4$ = 0.133.

***Construction of the target function for the 2D spectra.*** The simulations of the 2D spectra involves the five geometrical parameters $\theta,\phi,\alpha,\beta$ and $R$; the line-broadening parameters $\sigma_{RP}$ , $\sigma_{NRP}$, $\kappa_{RP}$ and $\kappa_{NRP}$ discussed above; and the doubly-excited state manifold fluorescence efficiency parameter Γ. For the least-square analysis of 2D spectra we used the experimental data in the frequency window $\omega_\tau \in$ [3.04 rad fs$^{-1}$, 3.15 rad fs$^{-1}$] and $\omega_t \in$ [3.04 rad fs$^{-1}$, 3.15 rad fs$^{-1}$], where the most intense diagonal peaks and cross-peaks were located. The least-square sum $\chi^2_{2D}$ includes the six sets of 2D experimental data, i.e., the real, imaginary and absolute value spectra for rephasing and non-rephasing signals. For example, the value of $\chi^2_{2D}$ for the best fit to both linear and 2D spectra is 9.87.

***Importance of the combined target function.*** Finding a single conformation that agrees well with the linear and 2D data proved to be a restrictive task, suggesting a definitive structural determination. For example, the optimization of either $\chi^2_{lin}$ or $\chi^2_{2D}$ by themselves did not result in solutions that were consistent with the other type of spectra. A single solution was only possible when the combined target function $\chi^2_{tot} = \chi^2_{lin} + \chi^2_{2D}$ was used. As shown in Fig. 4 of the text, it was possible to find examples for which $\chi^2_{lin}$ was smaller than the value obtained for the optimal conformation. Yet in these cases the 2D spectra departed significantly from the experimental data. Similarly, the optimization of only the target function $\chi^2_{2D}$ could lead to misleading results. In Table S1, we list values for the target function and its linear and 2D components for several values of the structural angles, which were scanned relative to the optimized conformation. We note that Table S1 contains some negative values for either $\chi^2_{lin}$ or $\chi^2_{2D}$, indicating that a departure from the $\chi^2_{tot}$ minimum can yield improved agreement with one type of spectra at the expense of agreement with the other. The results presented in Table S1 suggests that the sensitivity of the search to structural parameters allows for a quantitative estimate of dimer conformation.

**6. Error Analysis and Propagation of Uncertainties in PM-2D FS Signals.** In this section we calculate trust intervals for the structural parameter values we have obtained for the MgTPP dimers embedded in DSPC liposomes. We discuss here the uncertainties in our results, which arise from two different sources: 1) the quality of the optimization search performed with the KNITRO package, and 2) the uncertainty in the reference experimental data used to construct the target function $\chi^2_{tot}$.

To determine the quality of the KNITRO search, e.g., the absence of convergence to local minima, we performed a fine-resolution parameter scan to verify the extent to which the values obtained by the program indeed correspond to a global minimum of the target function, i.e., the best minimum from the multi-start search. In Fig. S3, we plot the relative deviation $\Delta\chi^2_{tot}/\chi^2_{tot} = \left(\chi^2 - \chi^2_{tot,ref}\right)/\chi^2_{tot,ref}$ from the reference value of $\chi^2_{tot,ref}$, which can be interpreted as a relative error when moving away from the optimal conformation. Fig. S3 shows that the structure found is the minimum, to within ± 1° for the each of the angles, ± 0.05 Å for the $R$ distance, and ± 0.01 units in Γ. The few missing points in the scans for $\alpha$ and $\phi$ were removed because these converged to a higher local minima above the predominant-branch where the majority of points appear to lie. For all of the scans, one parameter was varied while the remaining parameters that entered the calculation of the 2D spectra were held constant. The lack of convergence we refer to here is due to the additional optimization required to relax the parameters needed for the linear spectra (i.e., $a_0$ , $\{a_i\}$, η and $\sigma_{lin}$ in Eq. S22). Since the few data points that converged above the predominant-branch do not suggest an alternative minimum, it was not necessary to converge these points since enough were present to clearly show the behavior upon approaching the minimum.

The scans in Fig. S3 also serve to assess the degree of sensitivity. For example, it is clear that the scans are more sensitive to the parameters $\beta$, $R$, and $\theta$, when compared to other degrees of freedom such as $\alpha$, $\phi$, and Γ. As a consequence, under a certain fixed relative error, one expects that the uncertainty will be smaller for $\beta$ and $\theta$ while slightly larger for $\alpha$ and $\phi$.

Having established that our search routine is almost exact, we next address the error propagation due to uncertainties in the experimental measurements. In the following, we base our discussion on $\chi^2_{2D}$ motivated by the assumption that $\Delta\chi^2_{tot}/\chi^2_{tot} \approx \Delta\chi^2_{2D}/\chi^2_{2D}$, i.e., that these relative errors are comparable. We thus use our estimate of $\Delta\chi^2_{2D}/\chi^2_{2D}$ to read out the trust intervals directly from the scans shown in Fig. S3. This relative error was estimated to be approximately 1%, and it is indicated separately for each structural parameter by the red-shaded rectangles in Fig. S3.

We next explain the assumptions we have made to obtain the 1% estimate using standard error propagation analysis (10). The 2D target function is defined according to

$$
\begin{aligned}
\chi^2_{2D} = \sum_{\omega_\tau^i,\omega_t^j} & \left\{\mathrm{Abs}\left[\mathrm{NRP}_{sim}\left(\omega_\tau^i,\omega_t^j\right)\right] - \mathrm{Abs}\left[\mathrm{NRP}_{exp}\left(\omega_\tau^i,\omega_t^j\right)\right]\right\}^2 \\
& + \left\{\mathrm{Re}\left[\mathrm{NRP}_{sim}\left(\omega_\tau^i,\omega_t^j\right)\right] - \mathrm{Re}\left[\mathrm{NRP}_{exp}\left(\omega_\tau^i,\omega_t^j\right)\right]\right\}^2 \\
& + \left\{\mathrm{Im}\left[\mathrm{NRP}_{sim}\left(\omega_\tau^i,\omega_t^j\right)\right] - \mathrm{Im}\left[\mathrm{NRP}_{exp}\left(\omega_\tau^i,\omega_t^j\right)\right]\right\}^2 \\
& + \left\{\mathrm{Abs}\left[\mathrm{RP}_{sim}\left(\omega_\tau^i,\omega_t^j\right)\right] - \mathrm{Abs}\left[\mathrm{RP}_{exp}\left(\omega_\tau^i,\omega_t^j\right)\right]\right\}^2 \\
& + \left\{\mathrm{Re}\left[\mathrm{RP}_{sim}\left(\omega_\tau^i,\omega_t^j\right)\right] - \mathrm{Re}\left[\mathrm{RP}_{exp}\left(\omega_\tau^i,\omega_t^j\right)\right]\right\}^2 \\
& + \left\{\mathrm{Im}\left[\mathrm{RP}_{sim}\left(\omega_\tau^i,\omega_t^j\right)\right] - \mathrm{Im}\left[\mathrm{RP}_{exp}\left(\omega_\tau^i,\omega_t^j\right)\right]\right\}^2 .
\end{aligned}
\qquad \text{(S23)}
$$

In Eq. S23, the subscripts "*sim*" and "*exp*" indicate simulated and experimental spectra, respectively. The indices "*i*" and "*j*" indicate the 2D frequency coordinate. For the error propagation analysis, we include every data point from each of the six Fourier-transformed experimental signals [Abs(NRP*exp*), Abs(RP*exp*), Re(NRP*exp*), Re(RP*exp*), Im(NRP*exp*), and Im (RP*exp*)] to define a variable with its own uncertainty. For simplicity, we define

$$\mathrm{Abs}\left[\mathrm{NRP}_{exp}\left(\omega_\tau^i,\omega_t^j\right)\right]\equiv f_1^{ij},\ \mathrm{Re}\left[\mathrm{NRP}_{exp}\left(\omega_\tau^i,\omega_t^j\right)\right]\equiv f_2^{ij},\ \mathrm{Im}\left[\mathrm{NRP}_{exp}\left(\omega_\tau^i,\omega_t^j\right)\right]\equiv f_3^{ij},$$

$$\mathrm{Abs}\left[\mathrm{RP}_{exp}\left(\omega_\tau^i,\omega_t^j\right)\right]\equiv f_4^{ij},\ \mathrm{Re}\left[\mathrm{RP}_{exp}\left(\omega_\tau^i,\omega_t^j\right)\right]\equiv f_5^{ij},\ \text{and}\ \mathrm{Im}\left[\mathrm{RP}_{exp}\left(\omega_\tau^i,\omega_t^j\right)\right]\equiv f_6^{ij}.$$

The sum in Eq. S23 is performed over the discrete frequency values inside the interval $\omega_\tau,\omega_t \in$ (3.04, 3.15) rad fs$^{-1}$. Since there are $N = 101$ data points per frequency axis inside this interval, the number of terms in the summation contains $N^2 = 10{,}201$ variables of the form $f_k^{ij}$ for each value of $k$. Since we are dealing with $k = 1 - 6$, the number of independent variables in the error propagation analysis is 61 206. We define $z \equiv \chi_{2D}^2\left(\left\{f_k^{ij}\right\}\right) = \chi_{2D}^2\left(\left\{g_n\right\}\right)$, where $g_n = f_k^{ij}$, with $n$ running from 1 - 61,206 denoting all possible combinations of $i$, $j$, and $k$. Under the assumption that all variables are independent, we estimate the uncertainty of $z$ by (10)

$$\Delta z = \sqrt{\sum_{n=1}^{61{,}206}\left(\frac{\partial z}{\partial g_n}\Delta g_n\right)^2}. \tag{S24}$$

In terms of the $g_n$ variable, Eq. S23 for $\chi_{2D}^2$ can be rewritten

$$z = \sum_{n=1}^{61{,}206}\left(g_n^{sim} - g_n\right)^2. \tag{S25}$$

The partial derivative can be calculated according to $\partial z/\partial g_n = -2\left(g_n^{sim} - g_n\right)^2$. Once the uncertainties $\Delta g_n$ are calculated, the error in Eq. S24 can be easily determined.

As previously stated, each of the $g_n$ corresponds to a data point from any of the 2D spectra involved in the calculation of $\chi_{2D}^2$. To estimate the uncertainty associated with each of the 61,206 variables, we divide them into two groups; the first half (*n* = 1 - 30,603) associated with the absolute value, real and imaginary parts of the rephasing data, and the remaining half (*n* = 30,604 - 61,206) associated with that of the non-rephasing data. To simplify these calculations, we find a single uncertainty value representative for each of the two types of spectra. We denote these as $\Delta g_{RP}$ and $\Delta g_{NRP}$ for the rephasing and non-rephasing data, respectively. Calculations of these uncertainties are illustrated in Fig. S4. The uncertainty is estimated from four different experimental runs performed on a ZnTPP monomer in dimethylformamide solution, which were processed using an identical procedure to the MgTPP samples studied here. The 2D absolute value rephasing and non-rephasing spectra of one data run are shown in Figs. S4 A and S4 B, respectively. In Figs. S4 C and S4 D are shown overlays of the absolute value rephasing and non-rephasing signals, $s_\omega^{RP(NRP)}$, for each of the four data runs along the diagonal profile, with $\omega_\tau = \omega_t = \omega$. Figs. S4 E and S4 F show the average signal $\bar{s}_\omega^{RP(NRP)} \equiv \left\langle s_\omega^{RP(NRP)} \right\rangle_{sets}$ along the diagonal profile, where $\langle \cdots \rangle_{sets}$ indicates the average performed over individual data sets. We similarly calculate the variance at each value of $\omega$ according to $\sigma_{RP(NRP)}^2(\omega) = \left\langle \left[ s_\omega^{RP(NRP)} - \bar{s}_\omega^{RP(NRP)} \right]^2 \right\rangle_{sets}$, which are shown Figs. S4 G and S4 H.

The representative uncertainties, $\Delta g_{RP}$ and $\Delta g_{NRP}$, are estimated as the frequency average of the standard deviations along the diagonal profiles, i.e., $\Delta g_{RP(NRP)} = \left\langle \sigma_{RP(NRP)}(\omega) \right\rangle_\omega$. The average over frequency was done to include most of the significant data, taking approximately twice the full-width at half-maximum from the main peak for both the rephasing and non-rephasing profiles - i.e., over the interval $\omega \in$ (3.07, 3.20) rad fs$^{-1}$. By using the resulting values for $\Delta g_{RP}$ = 0.0086 and $\Delta g_{NRP}$ = 0.016 in Eq. S24, we find that $\Delta z / z_{ref} = \Delta\chi^2_{2D} / \chi^2_{2D} \approx \Delta\chi^2_{tot} / \chi^2_{tot}$ = 0.0096 ~ 1%. The value of $\chi^2_{2D}$ = 9.87 used for this estimate corresponds to the reference value obtained for the optimal conformation. Having established that the expected error is ~ 1%, we determine the trust intervals directly from the parameter scan plots shown in Fig. S3, as indicated by the red-shaded rectangles. These intervals correspond to -16° < $\Delta\theta$ < 4°, -11° < $\Delta\phi$ < 11°, -11° < $\Delta\alpha$ < 11°, -2° < $\Delta\beta$ < 2°, -0.05 Å < $\Delta R$ < 0.05 Å, and -0.1 < $\Delta\Gamma$ = 0.1, where $\Delta x \equiv x - x_{ref}$, and $x_{ref}$ is taken from the optimized outcomes.

We conclude this section by commenting on the uncertainty of the variable $R$. In addition to the uncertainties discussed above, an accurate estimate of $\Delta R$ must also account for its dependence on the calculated value of the monomer square transition dipole moment $|\mu|^2$. Uncertainty in the estimation of $|\mu|^2$ (Eq. S4) will appear in the electronic couplings (Eq. S3) as a rescaling of the end-to-end distance $R$. For example, too small an estimation of $|\mu|^2$ will result in an apparent value of $R$ that is also too small. Although we have attempted to make our estimate of $|\mu|^2$ as accurate as possible, we cannot discount the possibility that a systematic error is present. We note that the values we have obtained for the angles $\theta$ , $\phi$ , $\alpha$, and $\beta$ constrain the conformation significantly. We therefore propose that further refinements in the conformation could be achieved through quantum chemical calculations. For example, semi-empirical calculations on the MgTPP dimer, in which only the distance $R$ is varied, could be used to obtain its value where the energy minimum occurs. Given the degree of molecular detail provided by quantum chemical calculations, it should in principle be possible to capture the effects of steric interactions between bulky phenyl groups. Such an approach might be useful to further refine the values of the structural parameters within their trust intervals.

**References**

1. MacMillan JB, and T. F. Molinski (2004) Long-range stereo-relay: Relative and absolute configuration of 1,n-glycols from circular dichroism of liposomal porphyrin esters. *J. Am. Chem. Soc.* 126:9944-9945.
2. Tekavec PF, G. A. Lott, and A. H. Marcus (2007) Fluorescence-detected two-dimensional electronic coherence spectroscopy by acousto-optic phase modulation. *J. Chem. Phys.* 127:214307.
3. Gouterman M (1979) *The Porphyrins*, ed Dolphin D (Academic Press, New York), Vol III, pp 1-156.
4. Stomphorst RG, R. B. M. Koehorst, G. van der Zwan, B. Benthem, and T. J. Schaafsma (1999) Excitonic interactions in covalently linked porphyrin dimers with rotational freedom. *J. Porphyrins and Phthalocyanines* 3:346-354.
5. Koolhaas MHC, G. van der Zwan, F. van Mourik, and R. van Grondelle (1997) Spectroscopy and structure of bacteriochlorophyll dimers. I. Structural consequences of nonconservative circular dichroism spectra. *Biophys. J.* 72:1828-1841.
6. Won Y, R. A. Friesner, M. R. Johnson, and J. L. Sessler (1989) Exciton interactions in synthetic porphyrin dimers. *Photonsynthetic Research* 22(3):201-210.
7. Hardwick JL (2003) Absorption and emission of electromagnetic radiation. *Handbook of Molecular Physics and Quantum Chemistry*, ed Wilson S (Wiley).
8. Mukamel S (1995) *Nonlinear Optical Spectroscopy* (Oxford University Press, Oxford).
9. Cho M (2009) *Two-Dimensional Optical Spectroscopy* (CRC Press, Boca Raton) 1st Ed.
10. Taylor JR (1997) *An Introduction to Error Analysis: The Study of Uncertainties in Physical Measurements* 2nd Ed.

## Supporting Information Figure Legends

Figure S1. (A) Collinear sequence of optical pulses used in PM-2D FS experiments. The coherence, population, and measurement periods ($\tau, T,$ and $t$) are indicated, as well as the relative phase of pulses 1 and 2 ($\phi_{21}$), and pulses 3 and 4 ($\phi_{43}$). (B) Schematic of the PM-2D FS apparatus, described in the text and in (2). The phases of the pulse electric fields are swept using acouto-optic Bragg cells, which are placed in the arms of two Mach-Zehnder interferometers (MZI 1 & MZI 2). The excitation pulses are made to be collinear before entering the sample. Reference waveforms are constructed from the pulse pairs from each interferometer. The reference signals oscillate at the difference frequencies of the acousto-optic Bragg cells (5 kHz and 8 kHz for ref 1 and ref 2, respectively). The reference signals are sent to a waveform mixer to construct "sum" and "difference" side band signals (3 kHz and 13 kHz). These reference side bands are used to phase-synchronously detect the fluorescence, which isolates the non-rephasing and rephasing population terms, respectively.

Figure S2. Double-sided Feynman diagrams (DSFD) representing the light-matter interactions contributing to the rephasing and non-rephasing signals measured experimentally. The four-level model used to describe the coupled dimers of MgTPP are shown in Fig. 1*A* of the text. The collective dipole moment allows transitions from the ground state to the first-excited manifold, and from the latter to the final doubly-excited state. The sign associated with each diagram is determined by the number of arrows (dipole interactions) on the right vertical line of each ladder diagram ("bra" side). An even (odd) number of interactions picks up a positive (negative) sign for the term under consideration. Therefore, the non-rephasing and rephasing 2D PE signals are $S_{NRP}^{2D\,PE}(\tau,T,t) \propto \mathbf{R}_{1a}^{*} + \mathbf{R}_{2a} - \mathbf{R}_{3b}^{*}$ and $S_{RP}^{2D\,PE}(\tau,T,t) \propto \mathbf{R}_{4a} + \mathbf{R}_{3a} - \mathbf{R}_{2b}^{*}$, respectively, while the corresponding PM-2D FS signals are $S_{NRP}^{PM-2D\,FS}(\tau,T,t) \propto -\left(\mathbf{Q}_{5a}^{*} + \mathbf{Q}_{2a} + \mathbf{Q}_{3b}^{*} - \Gamma\mathbf{Q}_{7B}\right)$ and $S_{RP}^{PM-2D\,FS}(\tau,T,t) \propto -\left(\mathbf{Q}_{4a} + \mathbf{Q}_{3a} + \mathbf{Q}_{2b}^{*} - \Gamma\mathbf{Q}_{2b}^{*}\right)$. The parameter $\Gamma$ accounts for the different fluorescence quantum yields between doubly- and singly-excited state manifolds.

Figure S3. Relative deviation of the target function, $\Delta\chi_{tot}^{2}/\chi_{tot}^{2}$ , from the optimized reference value, $\chi_{tot,ref}^{2}$ , as a function of structural parameter uncertainties. Cross-sections of the target function are shown for the uncertainties (A) $\Delta\theta$, (B) $\Delta\phi$, (C) $\Delta\alpha$ , (D) $\Delta\beta$ , (E) ΔR, and (F) ΔΓ, where $\Delta x \equiv x - x_{ref}$, and $x_{ref}$ is the value corresponding to the optimized conformation. The optimized conformation corresponds to a minimum of the multi-dimensional parameter surface. As indicated by the red shaded rectangles, trust intervals are directly read out from these plots, based on the ~ 1% relative error associated with the experimental data quality. The trust interval regions are expanded and shown as insets for the parameters $\Delta\theta$ , $\Delta\beta$ , and ΔR. The resulting intervals are -16° < $\Delta\theta$ < 4°, -11° < $\Delta\phi$ < 11°, -11° < $\Delta\alpha$ < 11°, -2° < $\Delta\beta$ < 2°, -0.05 Å < $\Delta R$ < 0.05 Å, and -0.1 < ΔΓ = 0.1.

Figure S4. Experimental data runs performed on ZnTPP monomer in dimethylformamide solution, which were used for error propagation analysis. In panels (A) and (B) are shown, respectively, the 2D absolute value rephasing and non-rephasing spectra of a single representative data set. In panels (C) and (D) are shown overlays of the absolute value rephasing and non-rephasing signals for each of the four data runs along the diagonal profile. Panels (E) and (F) show the average of the four data sets along the diagonal profile. In panels (G) and (H) are shown the corresponding variances along the diagonal profile. By integrating the standard deviation of the data over the interval $\omega \in$ (3.07, 3.20) rad fs$^{-1}$, we obtain the relative uncertainties $\Delta g_{RP}$ = 0.0086 and $\Delta g_{NRP}$ = 0.016 (defined in SI text). These values are input to Eq. S24 to estimate the relative target function uncertainty $\Delta\chi^2_{tot}/\chi^2_{tot}$ = 0.0096 ≃ 1%, which in turn establishes the trust intervals of the structural parameters relative to the optimized outcome.

**Supporting Information Table Legends**

Table S1. Linear least-square target function $\chi^2_{tot} = \chi^2_{lin} + \chi^2_{2D}$ dependence on structural angles. Target function values are given relative to the reference values: $\chi^2_{lin}$ = 7.39, $\chi^2_{2D}$ = 9.87, and $\chi^2_{tot}$ = 17.26, which correspond to the conformation with structural parameters $\theta$ = 117.4°, $\phi$ = 225.2°, $\alpha$ = 135.2°, $\beta$ = 137.2°, $R$ = 4.2 Å, and Γ = 0.31, and line-broadening parameters $\sigma_{RP}$ = 108.1 fs, $\sigma_{NRP}$ = 96.2 fs, $\kappa_{RP}$ = 98.1 fs, and $\kappa_{NRP}$ = 102.9 fs.

Table S2. Linear least-square target function $\chi^2_{2D}$ dependence on fluorescence efficiency Γ of the doubly-excited state manifold. Values are given relative to the optimized conformation with $\chi^2_{2D}$ = 9.87 and Γ = 0.31.

Figure S1.

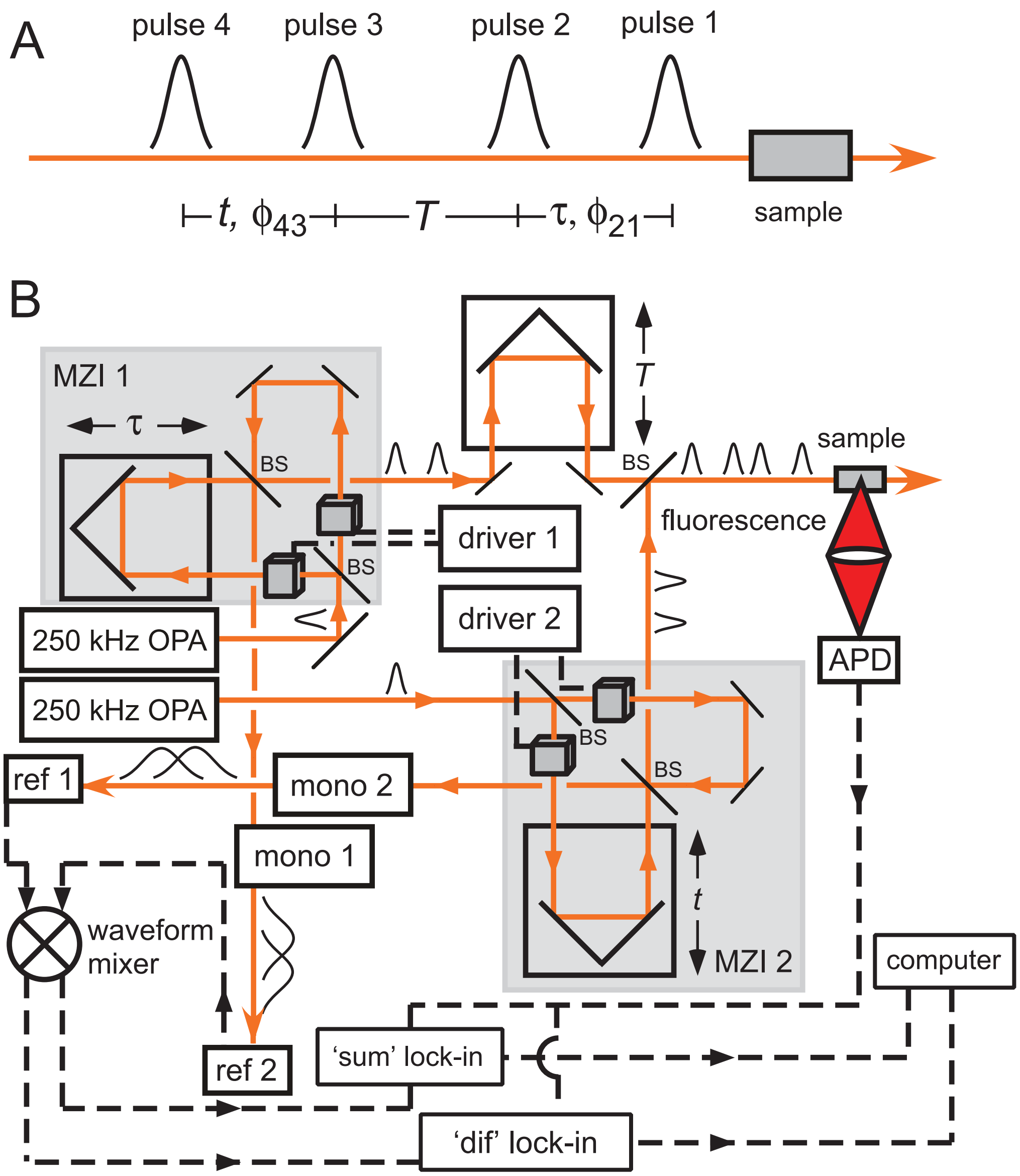

Figure S2.

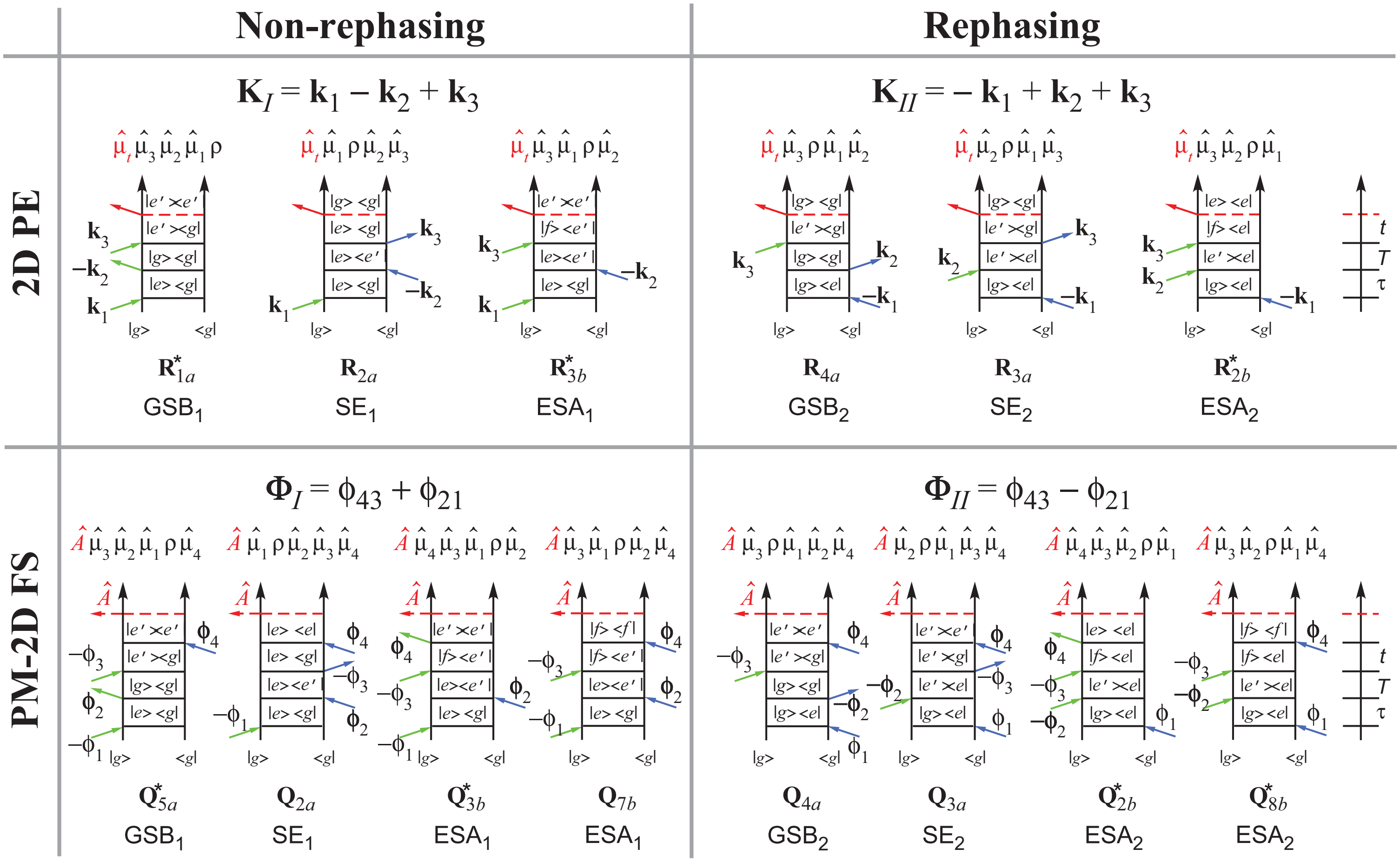

Figure S3.

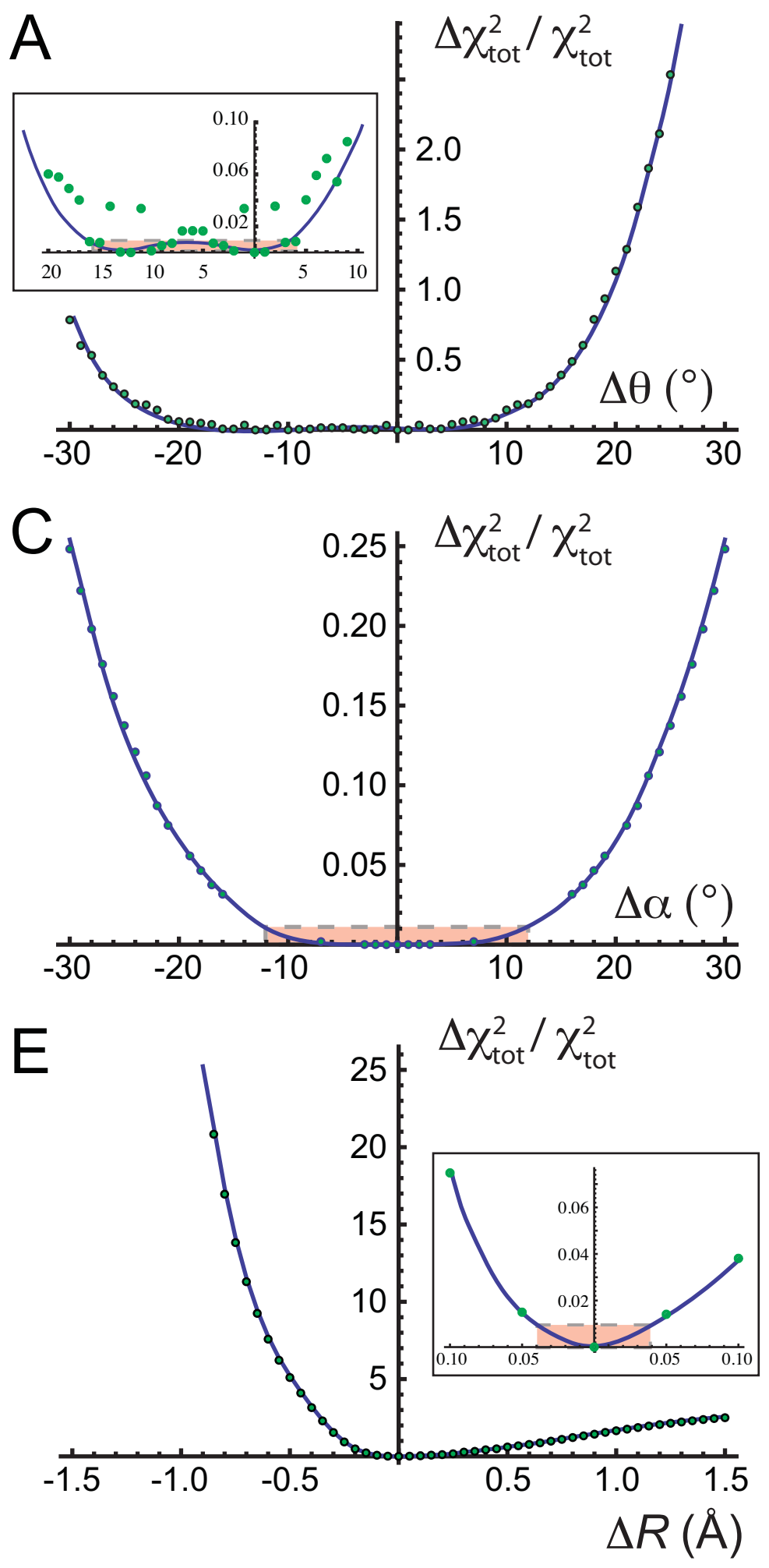

Figure S4.

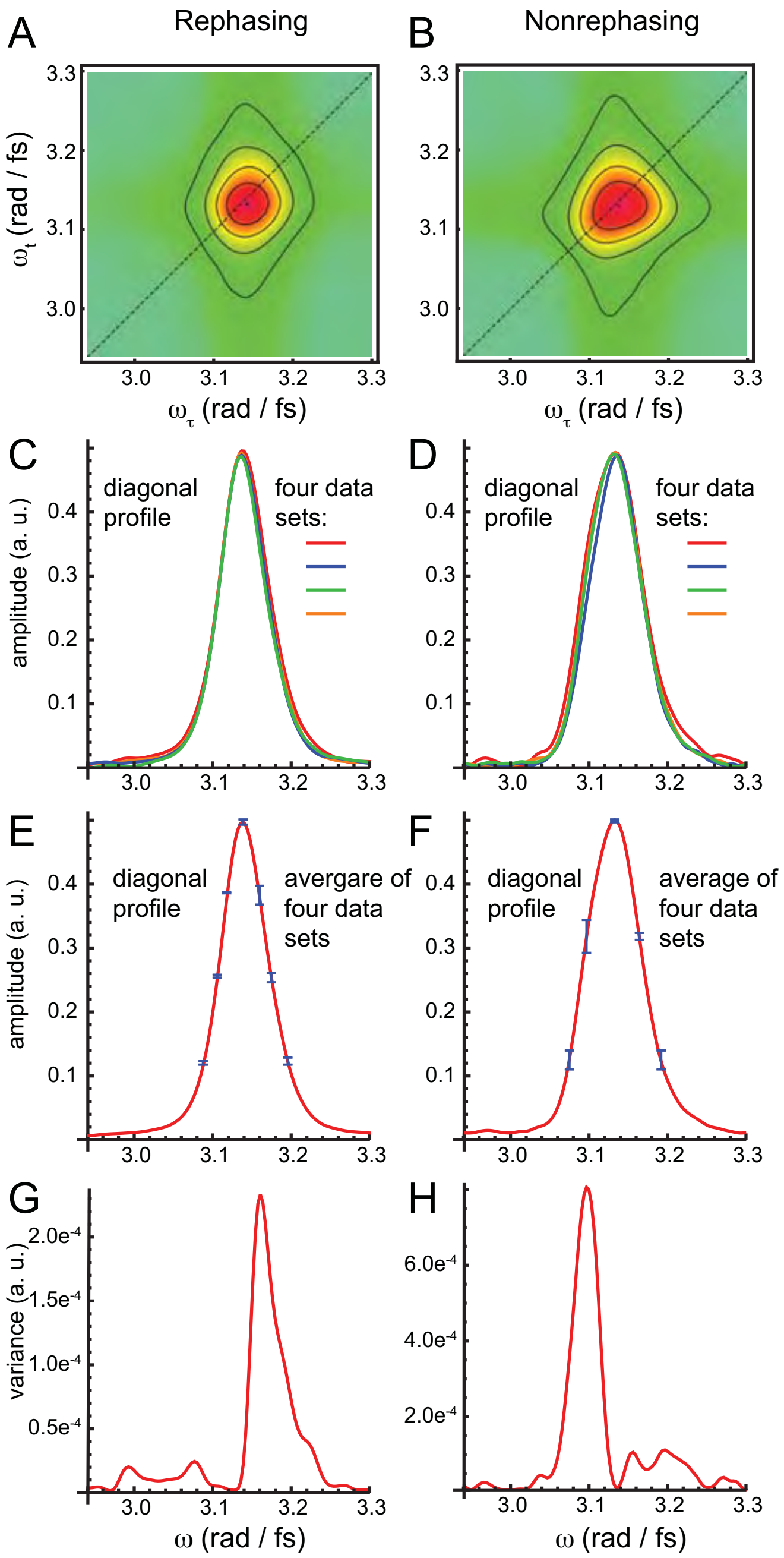

A
Rephasing
B
Nonrephasing
ω_t (rad / fs)
ω_τ (rad / fs)
C
diagonal profile
four data sets:
D
diagonal profile
four data sets:
amplitude (a. u.)
E
diagonal profile
avergare of four data sets
F
diagonal profile
average of four data sets
G
variance (a. u.)
H
ω (rad / fs)

Table S1.

| | $\Delta\theta$ | | | $\Delta\phi$ | | | $\Delta\alpha$ | | | $\Delta\beta$ | | |
|---|---|---|---|---|---|---|---|---|---|---|---|---|
| deg | $\Delta\chi^2_{lin}$ | $\Delta\chi^2_{2D}$ | $\Delta\chi^2_{tot}$ | $\Delta\chi^2_{lin}$ | $\Delta\chi^2_{2D}$ | $\Delta\chi^2_{tot}$ | $\Delta\chi^2_{lin}$ | $\Delta\chi^2_{2D}$ | $\Delta\chi^2_{tot}$ | $\Delta\chi^2_{lin}$ | $\Delta\chi^2_{2D}$ | $\Delta\chi^2_{tot}$ |
| -30 | 8.12 | 5.45 | 13.6 | 3.58 | 0.81 | 4.39 | 3.58 | 0.81 | 4.39 | 141 | -0.35 | 141 |
| -25 | 2.31 | 2.14 | 4.45 | 1.93 | 0.5 | 2.43 | 1.93 | 0.5 | 2.43 | 101 | -0.34 | 101 |
| -20 | 0.33 | 0.72 | 1.04 | 0.97 | 0.28 | 1.25 | 0.97 | 0.28 | 1.25 | 64.7 | -0.31 | 64.4 |
| -15 | -0.02 | 0.15 | 0.13 | 1.08 | 0.14 | 1.22 | 1.08 | 0.14 | 1.22 | 34.8 | -0.26 | 34.6 |
| -10 | 0.08 | -0.06 | 0.02 | 0.61 | 0.06 | 0.67 | 0.61 | 0.06 | 0.67 | 14.1 | -0.19 | 13.9 |
| -5 | 0.38 | -0.1 | 0.29 | 0.56 | 0.01 | 0.57 | 0.56 | 0.01 | 0.57 | 3.03 | -0.11 | 2.92 |
| 0 | 0 | 0 | 0 | 0 | 0 | 0 | 0 | 0 | 0 | 0 | 0 | 0 |
| 5 | 0.38 | 0.32 | 0.7 | 0.56 | 0.01 | 0.57 | 0.56 | 0.01 | 0.57 | 1.03 | 0.15 | 1.18 |
| 10 | 1.33 | 1.15 | 2.48 | 0.61 | 0.06 | 0.67 | 0.61 | 0.06 | 0.67 | 3.22 | 0.36 | 3.58 |
| 15 | 3.6 | 3.17 | 6.77 | 1.08 | 0.14 | 1.22 | 1.08 | 0.14 | 1.22 | 5.53 | 0.68 | 6.21 |
| 20 | 11.9 | 7.67 | 19.6 | 0.97 | 0.28 | 1.25 | 0.97 | 0.28 | 1.25 | 7.35 | 1.19 | 8.54 |
| 25 | 27.2 | 16.6 | 43.8 | 1.93 | 0.5 | 2.43 | 1.93 | 0.5 | 2.43 | 7.88 | 2.01 | 9.88 |
| 30 | 52.4 | 32.7 | 85 | 3.58 | 0.81 | 4.39 | 3.58 | 0.81 | 4.39 | 7.00 | 3.33 | 10.3 |

Table S2.

| $\Gamma$ | 0 | 0.2 | 0.31 | 0.4 | 0.6 | 0.8 | 1.0 | 1.2 | 1.4 | 1.6 | 1.8 | 2.0 |
|---|---|---|---|---|---|---|---|---|---|---|---|---|
| $\Delta\chi^2_{2D}$ | 0.857 | 0.104 | 0 | 0.083 | 0.814 | 2.319 | 4.580 | 7.520 | 10.87 | 13.97 | 17.25 | 21.20 |